\documentstyle[12pt]{article}

\renewcommand{\theequation}{\arabic{section}.\arabic{equation}}
\newcommand{\be}{\begin{equation}}
\newcommand{\ee}{\end{equation}}
\newcommand{\bea}{\begin{eqnarray}}
\newcommand{\eea}{\end{eqnarray}}

\def\tr{{\mathrm Tr\,}}                  %
\def\tr{{\mathrm Tr\,}}                  %
\def\Ad{{\mathrm Ad\,}}                  %
\def\ad{{\mathrm ad }}                   %
\def\adR{{\mathrm ad}_r}                 %
\def\pa{\partial}                        %
\def\wt{\widetilde}                      %
\def\G{{\cal G}}                         %
\def\Z{{\bf Z}}                          %
\def\M{{\cal M}}                         %
\def\H{{\cal H}}                         %
\def\A{{\cal A}}                         %
\def\F{{\cal F}}                         %
\def\P{{\cal P}}                         %
\def\E{{\cal E}}                         %
\def\T{{\cal T}}                         %
\def\S{{\cal S}}                         %
\def\P{{\cal P}}                         %
\def\cR{{\cal R}}                        %
\def\R{{\bf R}}                          %
\begin{document}

\thispagestyle{empty}
\setcounter{page}{0}

\renewcommand{\thefootnote}{\fnsymbol{footnote}}
\fnsymbol{footnote}

\vspace{.4in}

\begin{center} \Large\bf 
Chiral Extensions of the WZNW Phase Space,\\

Poisson-Lie Symmetries and Groupoids
\end{center}

\vspace{.1in}

\begin{center}
J.~Balog$^{(a)}$, L.~Feh\'er$^{(b)}$\protect\footnote{
Corresponding author's e-mail: lfeher@sol.cc.u-szeged.hu}
and L.~Palla$^{(c)}$ \\

\vspace{0.2in}

$^{(a)}${\em  Research Institute for Nuclear and Particle Physics,} \\
       {\em Hungarian Academy of Sciences,} \\
       {\em  H-1525 Budapest 114, P.O.B. 49, Hungary}\\

\vspace{0.2in}       

$^{(b)}${\em Institute for Theoretical Physics,
        J\'ozsef Attila University,} \\
       {\em  H-6726 Szeged, Tisza Lajos krt 84-86, Hungary }\\

\vspace{0.2in}   

$^{(c)}${\em  Institute for Theoretical Physics, 
Roland E\"otv\"os University,} \\
{\em  H-1117, Budapest, P\'azm\'any P. s\'et\'any 1 A-\'ep,  Hungary }\\

\end{center}

\vspace{.2in}

\begin{center} \bf Abstract
\end{center}

{\parindent=25pt
\narrower\smallskip\noindent
The chiral WZNW symplectic form $\Omega^{\rho}_{chir}$ is inverted 
in the general case. Thereby a precise relationship between the 
arbitrary monodromy dependent 2-form appearing in $\Omega^{\rho}_{chir}$  
and the exchange r-matrix that governs the  Poisson brackets of the 
group valued chiral fields is established. The exchange r-matrices are 
shown to satisfy a new dynamical generalization of the classical modified 
Yang-Baxter (YB) equation and Poisson-Lie (PL) groupoids are constructed 
that encode this equation analogously  as PL groups encode the classical 
YB equation. For an arbitrary  simple Lie group $G$, exchange r-matrices 
are found that are in one-to-one correspondence with the possible PL 
structures on $G$ and admit them as PL symmetries.
}

\vspace{11 mm}
\begin{center}
keywords: WZNW model, exchange algebra, Poisson-Lie symmetry 
\end{center}
\vfill\eject

\section{Introduction}
\setcounter{equation}{0}
\renewcommand{\thefootnote}{\arabic{footnote}}
\setcounter{footnote}{0}

This paper contains a systematic analysis 
of the classical phase space that arises from the chiral separation 
of the degrees of freedom in the Wess-Zumino-Novikov-Witten (WZNW) 
model \cite{Witt}.
The WZNW model occupies  a central position in conformal field 
theory \cite{DiF}.
Various structures that emerged from its 
study play an increasingly important r\^ ole in other areas 
of theoretical physics and in mathematics as well \cite{EFV}.
Among these structures are the quadratic exchange algebras that encode 
the Poisson brackets (PBs) of the chiral group valued fields, $g_C(x_C)$ for
$C=L,R$, which yield the general solution 
of the WZNW field equation as $g(x_L,x_R)= g_L(x_L) g^{-1}_R(x_R)$.
These exchange algebras were investigated intensively at the beginning 
of the decade (\cite{Bab} -- \cite{AT}) 
motivated  by the idea to understand the quantum group 
properties of the WZNW model \cite{qgroup} directly by means of canonical 
quantization \cite{Chu2,FHT,CL}. 
In accordance with the general philosophy of quantum groups \cite{Dri}, 
the Poisson-Lie (PL) symmetries of the chiral fields should be the most 
relevant in this respect.

The chiral WZNW Poisson structures found in the literature have the form 
\be 
\Big\{g_C(x)\stackrel{\otimes}{,} g_C(y)\Big\}
={1\over\kappa_C}\Big(g_C(x)\otimes
g_C(y)\Big)\Big(
\hat r + {1\over 2} {\hat I} \,{\mathrm sign}\,(y-x)
\Big), \quad 0< x,y<2\pi,
\label{1}\ee 
where $\hat I$ is given by the quadratic Casimir 
of the simple Lie algebra, $\G$,  of the WZNW group, $G$,  and the 
interesting object
is the `exchange r-matrix' $\hat r$. 
The choice of the PB is highly non-unique 
due to the fact that the $g_C$ are determined by the physical field $g$ 
only up to the gauge freedom $g_C\mapsto  g_C p$ for any constant $p\in G$.
In general $\hat r$ may depend on the monodromy matrix $M$ of the chiral 
fields, which satisfy $g_C(x+2\pi)= g_C(x) M$.
There are two qualitatively different cases that correspond to  
building the WZNW field out of chiral fields with diagonal 
monodromy (`Bloch waves')
or out of fields with generic monodromy.

For Bloch waves \cite{BDF,Chu5,BBT}, the Poisson structure is essentially 
unique and the associated r-matrix is a solution of the so called classical
dynamical Yang-Baxter (CDYB) equation, which has recently received a lot
of attention (see e.g.~the review in \cite{ES}).

For chiral fields with generic monodromy, 
it has been argued in \cite{Gaw,FG} that the possible exchange r-matrices 
should correspond to certain local differential 2-forms $\rho$
on open domains $\check G\subset G$, 
whose exterior derivative is the 3-form that occurs in the WZNW action. 
The precise connection between $\rho$ and $\hat r$ has not been elaborated, 
and in most papers dealing with generic monodromy actually 
only those very special cases were considered for which $\hat r$ is a 
monodromy independent constant.
In these cases $\hat r$ is necessarily a solution of 
the classical modified YB equation on the Lie algebra $\G$ 
with a certain definite normalization (eq.~(\ref{mCYBE}) with (\ref{hatf})).
This is a nice situation since if 
the same r-matrix is used to equip $G$ with a PL structure, then 
the gauge action of $G$ on the chiral WZNW field defines a PL symmetry. 
However, this mechanism of  PL symmetry is not available in
the physically most interesting case of a compact Lie group, 
because the relevant normalization admits no constant r-matrix for a 
compact $\G$.
Thus, in addition to the problem to understand the case of a general $\rho$,
an interesting question is whether the exchange r-matrix can be chosen
for a compact group in such a way to admit a PL symmetry on the 
chiral WZNW phase space. 

In this paper we study the family of chiral
exchange algebras (\ref{1}) concentrating on the case of generic monodromy
(for a related investigation of Bloch waves, see \cite{BFP2}).
Our main results are the following.

\begin{itemize}

\item
First,  we establish the relationship between the 2-form $\rho$ introduced in 
\cite{Gaw,FG}
and the corresponding  exchange r-matrix in the general case.
The result is given by eq.~(\ref{requfact}) with the notations in
(\ref{rqpm}), (\ref{qpara}), (\ref{rule}).

\item
Second, we point out a dynamical generalization  of the modified 
YB equation, eq.~(\ref{GCDYB}), whose solutions are 
the exchange r-matrices for generic monodromy.

\item
Third, we present explicitly a subfamily of the exchange r-matrices
whose members are in one-to-one correspondence with the possible PL structures 
on $G$ and admit them as PL symmetries. 
These exchange r-matrices, given by (\ref{standardcase}) with   (\ref{soln}),
contain the constant r-matrices studied earlier, and in another 
remarkable special case of them 
the gauge action of the group $G$ becomes a classical symmetry,
i.e., PL symmetry with the zero Poisson structure on $G$. 
They work for any (compact or not) simple Lie group.

\item
Fourth, we construct a family of PL groupoids that encode 
the dynamical YB equation (\ref{GCDYB}) analogously as PL groups encode 
the classical YB equation. This result generalizes a construction of 
\cite{EV} from diagonal to generic monodromy. 
 
\end{itemize}

The above mentioned results have been announced in \cite{BFP1} without proofs.
In addition to their detailed account, several other technical 
results can be found in this paper.  
The systematic exposition of the subject and the numerous examples 
that we present may be useful as a starting point for future studies.

The organization of the rest of the paper is as follows.
In the next section a necessary review of the chiral separation of the 
WZNW phase space is presented.
Section 3 contains a detailed account of the inversion of the possible 
symplectic structures on the chiral WZNW phase space, leading to the exchange
algebra (\ref{1}). 
Here many interesting additional issues are considered as well. 
In section 4 an alternative, shorter but less rigorous, 
derivation of the general exchange algebra is given, 
and a quick derivation of the exchange algebra of Bloch waves is also 
included.  Section 5 is devoted to  a general outline of the PL symmetries 
of the exchange algebra, and in particular to  the exchange r-matrices 
for which the standard  gauge action of $G$
on the chiral WZNW field yields such a symmetry.
Section 6 deals with the interpretation of the chiral WZNW Poisson structures
in terms of PL groupoids.
The paper ends with a discussion,
and  there are also two appendices containing some 
examples and the details of a proof.
 
\section{The WZNW solution space and its chiral extension}
\setcounter{equation}{0}

In this section we review the structure of the WZNW Hamiltonian system 
concentrating on the possible symplectic forms
on the chiral extension of its solution space, which are examined 
throughout the paper.
The presentation closely follows the line of thought found in \cite{FG}.

We consider a  simple, real or complex, Lie algebra, $\G$,  with a 
corresponding 
connected Lie group, $G$, and identify 
the phase space of the WZNW model associated with the group $G$ as 
\be
\M= T^* {\widetilde G}= \{\, (g,J_L)\,\vert\,
g\in \widetilde{G},\,\,\, J_L\in \widetilde{\G}\,\},
\label{WZphase1}\ee
where $\widetilde{G}=C^\infty(S^1,G)$ is the loop group and
$\widetilde{\G}=C^\infty(S^1,\G)$ is its Lie algebra.
The isomorphism of the  cotangent bundle $T^* {\widetilde G}$ with 
$\widetilde{G}\times \widetilde{\G}$ 
is established by means of right-translations on $\widetilde{G}$.
The elements $g\in \widetilde{G}$ (resp.~$J_L\in \widetilde{\G}$) are 
modeled as  $2\pi$-periodic
$G$-valued (resp.~$\G$-valued) functions on the real line ${\bf R}$.
The phase space is equipped with the symplectic form
\be
\Omega^\kappa= d \int_{0}^{2\pi} d\sigma\, 
{\mathrm Tr}\left( J_L d g g^{-1}\right)
+ {\kappa\over 2 }\int_{0}^{2\pi}d\sigma\,
 {\mathrm Tr}\left(d g g^{-1}\right) \wedge \left(d g g^{-1}\right)'
\label{WZsymp1}\ee
with some constant $\kappa$.
Here prime denotes derivative with respect to the space variable, 
$\sigma\in {\bf R}$,
and for any $A, B\in \G$ $\tr(AB)$ denotes a fixed multiple of the 
Cartan-Killing 
scalar product.
If $g$ and the `left-current' $J_L$ serve as coordinates on $\M$, 
then the `right-current' is given by    
\be
J_R=  - g^{-1} J_L g + \kappa g^{-1} g^\prime,
\label{I1}\ee
and in the alternative variables $(g,J_R)$ the symplectic form reads 
\be
\Omega^\kappa= -d \int_{0}^{2\pi} d\sigma\, {\mathrm Tr}
\left( J_R  g^{-1}dg \right)
- {\kappa\over 2 }\int_{0}^{2\pi}d\sigma\,
 {\mathrm Tr}\left(g^{-1}dg \right) \wedge \left(g^{-1}dg \right)'.
\label{IWZsymp2}\ee

Although the expression of $\Omega^\kappa$ appears rather formal at first 
sight,
it can be used to unambiguously associate hamiltonian vector fields 
and PBs
with a set of admissible functions, which include, for example, the 
Fourier components of 
$g$, $J_L$ and $J_R$. 
 We do not elaborate the precise meaning of the symplectic form here, 
since this is a standard matter in the context of the full WZNW model, 
but will face the analogous issue in the chiral context later, where it
is much less understood.
The only point that we wish to note  is that in the case of a complex 
Lie algebra
 the admissible functions depend holomorphically 
on the matrix elements of $g$, $J_L$, $J_R$ in the finite dimensional 
irreducible
representations of $G$, and $\widetilde{G}\times \widetilde{\G}$ is then 
a model of the holomorphic cotangent bundle.  

The phase space $\M$ represents the initial data for the WZNW system, whose
dynamics is generated by the Hamiltonian
\be
H_{\mathrm WZNW}={1\over 2\kappa} \int_0^{2\pi}d\sigma\, 
\tr\left(J_L^2 + J_R^2 \right).
\ee
Denoting time by $\tau$ and introducing lightcone coordinates as
\be
x_L := \sigma + \tau,
\quad x_R:=\sigma -\tau,
\quad
\pa_{L}= {\pa \over \pa x_L}= {1\over 2} ( \pa_\sigma + \pa_\tau),
\quad
\pa_R={\pa \over \pa x_R}= {1\over 2} ( \pa_\sigma - \pa_\tau),
\ee
Hamilton's equation can be written in the alternative forms \cite{Witt}
\be
\kappa\pa_L g = J_L g, \quad \pa_R J_L=0
\qquad{\Leftrightarrow}\qquad
\kappa\pa_R g = gJ_R, \quad \pa_L J_R=0.
\ee

Let $\M^{sol}$ be the space of solutions of the WZNW system.
$\M^{sol}$ consists of the smooth $G$-valued functions
$g(\sigma, \tau)$ which are $2\pi$-periodic in $\sigma$
and satisfy $\pa_R (\pa_L g\, g^{-1})=0$.
The general solution of this  evolution equation can be written as
\be
g(\sigma, \tau)= g_L(x_L) g_R^{-1}(x_R),
\label{gsol}\ee
where $(g_L, g_R)$ is any pair of $G$-valued, smooth,
quasiperiodic function on ${\bf R}$ with {\em equal monodromies}, i.e.,
for $C=L,R$ one has 
$g_C(x_C + 2\pi) = g_C(x_C)M$ 
with some $C$-independent  $M\in G$.
To elaborate this representation of the solutions in more detail,
we define the space $\widehat \M$:
\be
\widehat\M:= \{ (g_L, g_R) \vert g_{L,R} \in C^\infty({\bf R}, G),
\quad g_{L,R}(x + 2\pi) = g_{L,R}(x) M
\quad  M\in G\}.
\ee
There is a free right-action of $G$ on $\widehat \M$ given by
\be
G\ni p: (g_L, g_R) \mapsto (g_L p, g_R p).
\label{bundle}\ee
Notice that $\widehat \M$ is a {\em principal fibre bundle} over
$\M^{sol}$ with respect to the above action of $G$.
The projection of this bundle,
$\vartheta: \widehat\M \rightarrow \M^{sol}$, is given by
\be
\vartheta: (g_L, g_R)\mapsto g=g_L g_R^{-1}
\quad\hbox{i.e.}\quad g(\sigma,\tau)=g_L(x_L) g^{-1}_R(x_R).
\label{theta}\ee

We can identify $\M$ with $\M^{sol}$ by associating
the elements of the solution space with their initial data at $\tau=0$.
Formally, this  is described by the map
\be
\iota: \M^{sol} \rightarrow \M,
\qquad
\iota: \M^{sol}\ni  g(\sigma, \tau) \mapsto
\Bigl(g(\sigma, 0), J_L(\sigma)=(\kappa\pa_L g\, g^{-1})(\sigma,0)\Bigr)
\in \M.  \ee
Obviously,
$\iota^* (\Omega^\kappa)$ is then the natural symplectic form on 
the solution space.
Explicitly, 
\be
(\iota^* \Omega^\kappa)(g)=
-\kappa\left( d\,\int_{0}^{2\pi}d\sigma\,  
{\mathrm Tr}\left( g^{-1}\pa_R g\, g^{-1}dg \right)
+{1\over 2 }\int_0^{2\pi} d\sigma\,
 {\mathrm Tr}\left(g^{-1}dg \right) \wedge \pa_\sigma \left(g^{-1}dg \right)
\right)\big\vert_{\tau=0}
\label{solsymp}\ee
Regarding now $\M^{sol}$ as the base of the bundle
$\vartheta: \widehat{\M} \rightarrow \M^{sol}$, we obtain
a closed 2-form, $\widehat \Omega^\kappa$, on $\widehat\M$,
\be
\widehat \Omega^\kappa := \vartheta^* (\iota^* \Omega^\kappa) =
(\iota \circ \vartheta)^* \Omega^\kappa.
 \ee
By substituting the explicit formula (\ref{theta}) of $\vartheta$,
one finds that
\be
\widehat\Omega^\kappa(g_L, g_R)= \kappa_L \Omega_{chir}(g_L) +
\kappa_R \Omega_{chir}(g_R),
\label{*}\ee
where
\be\label{signdif}
\kappa_L := \kappa, \qquad
\kappa_R:= -\kappa,
\ee
and $\Omega_{chir}$ is the so called chiral WZNW 2-form:
\bea
&&\Omega_{chir}(g_C)=
- {1\over 2 }\int_0^{2\pi}dx_C\,
 {\mathrm Tr}\left(g_C^{-1}dg_C \right) \wedge \left(g_C^{-1}dg_C \right)'
-{1\over 2} \tr \left( (g_C^{-1} dg_C)(0)
\wedge dM_C 
{\scriptstyle\,} M_C^{-1}\right),\nonumber\\
&& M_C=g_C^{-1}(x) g_C(x+2\pi).
\label{**}\eea
This crucial formula of $\widehat\Omega^\kappa$ was first 
obtained by Gawedzki \cite{Gaw}.

It is clear from its definition 
that $d\widehat \Omega^\kappa=0$, but $\widehat \Omega^\kappa$ is not
a symplectic form on $\widehat{\M}$, since it is degenerate.
Of course,  its restriction
to any (local) section of the bundle
$\vartheta: \widehat{\M} \rightarrow \M^{sol}$ is a symplectic form,
since such sections yield (local) models of $\M^{sol}$.
On the other hand,  
one can check that  $\Omega_{chir}$ has a nonvanishing
exterior derivative \cite{Gaw}: 
\be
d\Omega_{chir}(g_C)= -{1\over 6} 
\tr\left( M_C^{-1}dM_C \wedge M_C^{-1}dM_C  \wedge 
M_C^{-1} dM_C \right).
\label{dOchir}\ee
Although this cancels from $d\widehat \Omega^\kappa$, since $M_L=M_R$ 
for the elements of $\widehat \M$,
it makes the chiral separation of the WZNW degrees of freedom a very
nontrivial and interesting problem.

The problem of the chiral separation can be described as follows \cite{FG}.
First, recall that the chiral currents $J_C$ ($C=L,R$)   
generate two commuting copies of the nontwisted affine Kac-Moody 
(KM) algebra 
of $\G$
and the WZNW field (\ref{gsol}) is a KM primary field  
under the Poisson bracket defined by the symplectic form on $\M^{sol}$.
In fact, by defining Fourier components as
\be
J_C^{\alpha,n}:= \int_0^{2\pi} dx_C\,e^{-inx_C} \tr(T^\alpha J_C)(x_C) 
\ee
using a basis\footnote{We have $I^{\alpha\beta}:=\tr(T^\alpha T^\beta)$ and
 $[T^\alpha, T^\beta]=f^{\alpha \beta}_\gamma
T^\gamma$ with summation over coinciding indices.}
 $T^\alpha$ of $\G$, 
it can be derived from (\ref{solsymp}) that the currents satisfy 
\be
\{ J_C^{\alpha,m}, J_C^{\beta,n}\} = 
f^{\alpha\beta}_\gamma J_C^{\gamma, m+n} 
+ 2i\pi \kappa_C m \delta_{m,-n} I^{\alpha\beta},
\qquad 
\{J_{L}^{\alpha,m}, J_R^{\beta,n}\}=0,
\label{KMPB}\ee
and
\be\label{Fourier}
\{ g(x_L, x_R), J_L^{\alpha, n}\} = e^{-in x_L} T^\alpha g(x_L, x_R),
\qquad
\{ g(x_L, x_R), J_R^{\alpha, n}\} = -e^{-in x_R} g(x_L, x_R)T^\alpha.
\ee
Second, the currents {\em almost} completely determine   
the chiral WZNW fields $g_{C}$, and thus also  $g=g_L g_R^{-1}$, 
by means of the differential equations
\be
\kappa_C \pa_C g_C = J_C  g_C \quad\hbox{for}\quad  C=L,R. 
\ee
Thus it appears an interesting possibility to construct the WZNW model 
as a reduction of a simpler model, in which the left and right-moving
degrees of freedom would be separated in terms of {\em completely 
independent} 
chiral  fields $g_L$ and $g_R$  
regarded as fundamental variables. 
It is clear that the solution space of such a chirally extended model 
must be 
a  direct product of two identical but independent spaces, i.e., it must 
have the form   
\be
\widehat\M^{ext} := \M_L \times \M_R 
\ee
with 
\be
\M_{C}:= \{ g_{C} \,\vert\, g_{C} \in C^\infty({\bf R}, G),
\quad g_{C}(x + 2\pi) = g_{C}(x) M_{C}
\quad  M_{C}\in G\}.
\ee
The space $\widehat \M^{ext}$ must be endowed with 
such a symplectic structure, $\widehat \Omega^\kappa_{ext}$, 
 that  reduces to 
$\widehat \Omega^\kappa$ on the submanifold 
$\widehat \M\subset \widehat\M^{ext}$ defined by the periodicity constraint 
\be
M_L=M_R.
\label{constraint}
\ee
It is easy to see that these requirements force 
$\widehat \Omega_{ext}^\kappa$
to have the following form:
\be
\widehat \Omega_{ext}^\kappa(g_L, g_R)= 
\kappa_L \Omega_{chir}^{\rho}(g_L) +\kappa_R \Omega_{chir}^{ \rho}(g_R)
\ee
where
\be
\Omega_{chir}^{\rho}(g_C) = \Omega_{chir}(g_C) + \rho(M_C) 
\label{Orho}\ee
with some 2-form $\rho$ depending {\em only} on the monodromy of $g_C$.
Since in  the extended model the factors  
$(\M_C, \kappa_C \Omega_{chir}^{\rho})$ must be {\em symplectic} 
manifolds {\em separately}, we have to satisfy the condition
\be
d\Omega_{chir}^{ \rho} = -{1\over 6} 
\tr\left( M_C^{-1} dM_C \wedge M_C^{-1}dM_C  \wedge 
M_C^{-1} dM_C \right) + d\rho(M_C) =0.
\label{dOrho}\ee
The problem now arises from the well-known fact that no globally 
defined smooth 2-form 
exists on $G$ that would satisfy this condition for all $M_C\in G$.  
 
There are two rather different wayouts from the above difficulty \cite{FG}.
The first is to restrict the possible domain of the monodromy 
matrix $M_C$ to some open submanifold in $G$ on which an appropriate 
2-form $\rho$ may be found. 
We refer to a choice of such a domain and a $\rho$ 
as {\em a chiral extension of the WZNW system},
and will explore the structure of the associated PB in the
subsequent sections.

The second possibility is to restrict the domain of the allowed monodromy
matrices much more drastically from the beginning, in such a way that after
the restriction $d\Omega_{chir}$ vanishes, whereby the difficulty disappears.
For example, one may 
achieve this by restricting the monodromy matrices to vary
in a fixed maximal torus of $G$, which amounts to constructing 
(a subset of) the 
solutions of the WZNW field equation in terms of chiral `Bloch waves'.
This second possibility is especially natural in the case of compact 
or complex Lie groups, for which there is only one maximal torus
up to conjugation.
The restriction to Bloch waves is equivalent to a partial (and local) gauge 
fixing of the bundle $\vartheta: \widehat \M \rightarrow \M^{sol}$.
The  resulting symplectic form is studied in detail in \cite{BFP2}.
 
\section{The chiral WZWN phase space}
\setcounter{equation}{0}

We here investigate the structure of the chiral WZNW phase space
$\M_C$ introduced in sec.~2.  
The analysis is the same for both chiralities, $C=L,R$, and we 
simplify our notation by putting $\M_{chir}$ 
for $\M_C$  and $g$, $M$, $J$, $\kappa$ for $g_C$, $M_C$, $J_C$, $\kappa_C$, 
respectively.   
We assume that the monodromy matrix $M$ is restricted  
to some open  submanifold $\check G \subset G$ on which a smooth 2-form 
$\rho$ is chosen in such a way that (\ref{dOrho}) holds. 
The domain in $\M_{chir}$ that corresponds to $M\in \check G$ is 
denoted by $\check \M_{chir}$. 
It turns out that $\kappa \Omega_{chir}^\rho$, 
defined by (\ref{Orho}) with (\ref{**}), is  
nondegenerate if  $\check G$ is appropriately chosen (so that eq.~(\ref{requ}) 
has a smooth, unique solution), 
and  we shall describe the general 
features of the PBs on $\check \M_{chir}$ associated with 
this symplectic form. We will then consider examples, in particular 
the  choices of $\rho$ introduced in \cite{FG}
that lead to Poisson-Lie symmetry on the full $\M_{chir}$.  

\subsection{Lie algebraic and differential geometric conventions}

Before we can turn to the task of inverting $\kappa \Omega_{chir}^\rho$,
we need to set up some conventions. 

An element $A\in \G$ has the components
$A_\alpha = \tr(AT_\alpha)$ and $A^\alpha=\tr(AT^\alpha)$
with respect to dual bases $T_\alpha$ and $T^\beta$ of $\G$:
\be
\tr(T_\alpha T^\beta)=\delta_\alpha^\beta,\qquad
I^{\alpha\beta}=\tr(T^\alpha T^\beta),
\qquad
I_{\alpha\beta}=\tr(T_\alpha T_\beta).
\ee
We will use $I^{\alpha\beta}$ and $I_{\alpha\beta}$ to raise and lower
Lie algebra indices.
Given a matrix $Q_{\alpha\beta}$, we can define an operator
$Q\in {\mathrm End}(\G)$ and an element $\hat Q\in \G\otimes \G$ by
\be
Q(A)= T_\alpha Q^{\alpha\beta} A_\beta,
\qquad
\hat Q= Q^{\alpha\beta} T_\alpha \otimes T_\beta.
\label{rule}\ee
The  matrix $I^{\alpha\beta}$ defines  the identity operator $I$, 
and $\hat I= T_\alpha \otimes T^\alpha$ is the `tensor Casimir'. 
For any $M\in G$, 
the matrix of the linear operator ${\mathrm Ad\,} M$ on $\G$,  
which we write as $({\Ad} M)(A)=M A M^{-1}$, 
will be denoted as 
\be
W_{\alpha\beta}(M)= \tr(T_\alpha M T_\beta M^{-1}).
\label{Wdef}\ee
The property $(\Ad M) (\Ad \tilde{M})=(\Ad M \tilde{M})$ 
yields the matrix multiplication rule 
\be
W_{\alpha\gamma}(M) I^{\gamma\theta} W_{\theta\beta}(\tilde{M}) = 
W_{\alpha\beta}(M \tilde{M}),
\ee
and we  also have 
\be 
W_{\beta\alpha}(M)=W_{\alpha\beta}(M^{-1})=
W^{-1}_{\alpha\beta}(M),
\qquad
W_{\alpha\beta}(M) I^{\beta\gamma} 
W^{-1}_{\gamma\theta}(M)=I_{\alpha\theta}.
\label{Wtilde}
\ee
Acting on a smooth function $\psi$ on $G$, we introduce the differential
operators 
\be
({\cal L}_\alpha \psi)(M):= {d\over d t}
 \psi(e^{t T_\alpha} M)\Big\vert_{t=0},
\qquad
 ({\cal R}_\alpha \psi)(M):= {d\over d t}
 \psi(Me^{t T_\alpha} )\Big\vert_{t=0},
\label{derLR}\ee
and their linear combinations
\be
{\cal D}_\alpha^\pm:= {\cal R}_\alpha \pm {\cal L}_\alpha.
\label{Dpm}\ee
For some purposes we will use a 
representation $\Lambda: G\rightarrow GL(V)$ of $G$ on a 
finite dimensional vector space $V$.
The corresponding representation of $\G$ is denoted by the same letter,
and we put 
\be
M^\Lambda:= \Lambda(M)
\quad\hbox{for}\quad M\in G,
\qquad  
T^\Lambda:= \Lambda(T)
\quad\hbox{for}\quad T\in \G.
\ee 
Such representations will be used (e.g.~the irreducible ones) that 
\be
\tr( T_\alpha T_\beta) = c_\Lambda 
{\mathrm tr\,}(T^\Lambda_\alpha T^\Lambda_\beta)
\label{indep}\ee 
holds with a constant $c_\Lambda$,
and we will then write 
\be
\tr(AB):= c_\Lambda {\mathrm tr\,}(AB),
\qquad
\forall A,B\in {\mathrm End}(V).
\ee
On the left hand side of eq.~(\ref{indep}) 
$\tr$ is a fixed (representation independent)
multiple of the Cartan-Killing form of $\G$, 
while $\mathrm tr$ is matrix trace over the representation space $V$. 

Remember that the phase space 
$\M_{chir}$ is parametrized by the $G$-valued field $g(x)$, 
which is assumed to be smooth in $x$ and is subject to the
monodromy condition 
\be 
g(x+2\pi)=g(x)M \qquad M\in G.  
\label{Moncon}\ee 
The corresponding chiral current, 
$J(x)=\kappa g^\prime(x)g^{-1}(x)\in{\cal G}$,
is thus a smooth, $2\pi$-periodic function of $x$. 
To define tangent vectors at
$g\in \M_{chir}$, we first have to consider 
smooth curves on the phase space.
Such a curve is given by a function
$\gamma(x,t)\in G$, which  is smooth in $x$,  $t$ 
and satisfies the deformed monodromy condition \be
\gamma(x+2\pi,t)=\gamma(x,t)M(t)\qquad
M(t)\in G.
\label{DefMon}\ee
To make sure that the curve passes through $g\in \M_{chir}$ 
at $t=0$,  we require 
$\gamma(x,0)=g(x)$.  
A vector $X[g]$ at $g\in \M_{chir}$
is obtained as the velocity to the curve at $t=0$,
encoded by the ${\cal G}$-valued, smooth function 
\be
\xi(x):={d\over dt} g^{-1}(x)\gamma(x,t)\Big\vert_{t=0}\,.
\label{Xidef}
\ee 
It is useful to note that, due to the analogous property of $g$,
the function $\xi$ on $\R$ may be reconstructed from
its restriction to $[0, 2\pi]$.
The monodromy properties of $\xi (x)$ can be derived by taking 
the derivative of (\ref{DefMon}),
\be 
\xi^\prime(x+2\pi)=M^{-1}\xi^\prime(x)M.
\label{XiMon}\ee 
This can be solved in terms of a smooth, $2\pi$-periodic 
$\G$-valued function, $X_J(x)$, and a constant Lie algebra element,
$\xi_0$, as follows: 
\be 
\xi(x)=\xi_0+\int_0^x dy\,
g^{-1}(y)X_J(y)g(y).
\label{XipJ}
\ee

A vector field $X$ on $\M_{chir}$ is an assignment,
$g\mapsto X[g]$,
of a vector to every point $g\in \M_{chir}$. 
It acts on a differentiable 
function, $g\mapsto F[g]$, on $\M_{chir}$ 
by the definition 
\be 
X(F)[g]={d\over d t}F[g_t]\Big\vert_{t=0}
\qquad g_t(x)=\gamma(x,t).  
\ee 
Since a  vector $X[g]$
can be parametrized by $\xi(x)$, which, in turn, can be parametrized
by the pair $\big(\xi_0,X_J(x)\big)$,
we can specify a vector field
by the assignments $g\mapsto \xi_0[g]\in \G$ and 
$g\mapsto X_J[g]\in\widetilde{\G}$.
Of course, the evaluation functions 
$F^x[g]:=g(x)$ and ${\cal F}^x[g]:=J(x)$ are differentiable with 
respect to any 
vector field, and their derivatives are given  by 
\be
X\big(g(x)\big)= g(x)\xi(x)
\qquad\hbox{and}\qquad
X\big(J(x)\big)=\kappa X_J(x),
\label{Xsigma}\ee 
which clarifies the meaning of $X_J$ as well. 
It is also obvious from (\ref{Moncon}) that the monodromy matrix 
yields a $G$-valued differentiable function on $\M_{chir}$, 
$g\mapsto M= g^{-1}(x) g(x+2\pi )$,
whose derivative is characterized by the $\G$-valued function 
\be
X(M) M^{-1}=  M\xi (x +2\pi) M^{-1} -\xi (x).
\label{XM}
\ee 

Having defined vector fields, one can now introduce differential 
forms as usual. 
We only remark that by (\ref{Xsigma}) evaluation 1-forms like 
$dg(x)$, $dJ(x)$ or $(g^{-1}dg)^\prime(x)$ are perfectly well-defined:
\bea
&& dg(x)\big( X\big)
=X(g(x))=g(x)\xi (x),\quad dJ(x)\big( X\big) =X(J(x))=\kappa
X_J(x),\nonumber\\
&& (g^{-1}(x)dg(x))^\prime \big( X\big) =\xi^\prime (x). 
\eea 
 
\subsection{Admissible Hamiltonians and  hamiltonian vector fields}

Now we turn to the following problem. For a fixed (scalar) function 
$F$ on the phase space $\check \M_{chir}$, 
we are looking for a corresponding vector field, $Y^F$, satisfying 
\be 
X(F)=\kappa\Omega^{\rho}_{chir}(X,Y^F) 
\label{hamvect}\ee 
for all vector fields $X$. 
Notice that $Y^F$ does not necessarily exist for a given $F$. 
We say that $F$ is an element of the set of {\em admissible Hamiltonians},
denoted as ${\tt H}$, 
if the corresponding hamiltonian vector field, $Y^F$,  exists.
Our purpose is to characterize ${\tt H}$ and to describe the mapping
${\tt H}\ni F \mapsto Y^F$.

We first compute $\kappa \Omega_{chir}^\rho(X,Y)$ for two vector 
fields $X$ and $Y$. 
Let $X$ be parametrized by $\xi(x)$ and further by the pair 
$\big(\xi_0,X_J(x)\big)$.  
The analogous parametrization for $Y$ is given  by $\eta(x)$ 
and the pair $\big(\eta_0,Y_J(x)\big)$.
Recall that $\kappa \Omega_{chir}^\rho$ is defined by  
eqs.~(\ref{**}), (\ref{Orho}), and parametrize $\rho$ now as 
\be 
\rho(M)= {1\over2}\,q^{\alpha\beta}(M){\mathrm Tr}
\big(T_\alpha M^{-1}dM\big)\wedge{\mathrm Tr}\big(T_\beta M^{-1}dM\big).
\label{qpara}\ee  
The $q^{\alpha\beta}$, $q^{\alpha\beta}=-q^{\beta\alpha}$, are smooth 
functions 
on the domain $\check G\subset G$, such that $d\Omega_{chir}^\rho=0$ 
on $\check \M_{chir}$.
A simple calculation, using partial integrations and (\ref{XM}),  
gives that 
\be
\kappa \Omega^{\rho}_{chir}(X,Y)=\kappa 
\Omega_{chir}(X,Y) + \kappa \rho(X,Y),
\label{Omega}\ee
where 
\bea
\Omega_{chir}(X,Y)&=&\int_0^{2\pi} dx 
{\tr}\Big(X_J(x)g(x)\Big(\eta(x)-{1\over2} M^{-1}Y(M)\Big) g^{-1}(x)\Big)
\nonumber\\
& - & {1\over2}{\tr} \left(\xi_0\Big(M^{-1} Y(M)+Y(M)M^{-1} \right)\Big)
\eea
and 
\be
\rho(X,Y) = 
\int_0^{2\pi} dx{\tr}\Big( X_J(x)g(x) B^Y(M) g^{-1}(x)\Big) 
 +{\tr}\Big(\xi_0\left(B^Y(M)-M B^Y(M) M^{-1}\right)\Big)
\ee
with 
\be
B^Y(M):=q^{\alpha\beta}(M)T_\alpha{\tr}\Big(T_\beta M^{-1} Y(M)\Big).  
\label{Hdef}\ee
Of course, all the expressions that appear in the above 
formula are functions 
of $g\in \check \M_{chir}$. 

Let us now suppose that $F\in {\tt H}$ and apply the above formula 
to $Y:=Y^F$.
Then the form of the right hand side of (\ref{hamvect}) implies
that there must exist a {\em smooth} $\G$-valued function on $\R$, 
$A^F(x)$, and a
constant Lie algebra element, $a^F$, such that for any vector field $X$ 
\be
X(F)=\kappa\int_0^{2\pi}dx \tr\Big(X_J(x)A^F(x)\Big)+ 
\kappa\tr\big(\xi_0a^F\big).
\label{Egy}\ee
This means that $F\in {\tt H}$ must have an exterior 
derivative parametrized
by the assignments $g\mapsto A^F(x)[g]$ and $g\mapsto a^F[g]$.
On the other hand, if $F$ is such that (\ref{Egy}) holds,   then 
we may try to solve (\ref{hamvect}) for the hamiltonian vector field.
Using the parametrization of $Y$ by $\eta(x)$,    
this leads to the following two equations: 
\be
\eta(x)-{1\over 2} M^{-1}Y(M) +B^{Y}(M) = g^{-1}(x)A^F(x)g(x),
\label{Elso}\ee
and 
\be
M^{-1}Y(M)+ Y(M) M^{-1} -2 B^{Y}(M) + 2 M B^{Y}(M) M^{-1} = -2a^F.
\label{Masodik} 
\ee
Now it is clear that $\eta^\prime(x)$ can be directly read 
off from (\ref{Elso}).
 {}From the identity $\eta'(x)=g^{-1}(x) Y_J(x) g(x)$, we then obtain that
\be
Y_J(x) = {A^F}^\prime(x)+{1\over\kappa}\Big[A^F(x),J(x)\Big].
\label{Ketto}
\ee 
This not only gives us the explicit formula of $Y_J$, but also means
that any $F\in {\tt H}$ must be such that the right hand side of (\ref{Ketto})
defines a {\em $2\pi$-periodic} smooth function of $x$.
Incidentally, this is equivalent to the monodromy condition (\ref{XiMon})
applied to the hamiltonian vector field $Y$.
To proceed further,  we use 
\be
Y(M) M^{-1}= M \eta(2\pi) M^{-1} - \eta(0).
\ee 
This implies that eqs.~(\ref{Elso}) and (\ref{Masodik}) 
are not linearly independent, and they can be
simultaneously solved for $\eta(x)$ only if one has 
\be
a^F=g^{-1}(0)\Big[A^F(0)-A^F(2\pi)\Big]g(0).
\label{Harom}\ee 
We now also see that the pair $(A^F, a^F)$ is uniquely determined 
for any $F\in {\tt H}$.
Indeed, the restriction of $A^F$ to $[0,2\pi]$ is completely fixed
by (\ref{Egy}), and is uniquely extended to a function on
$\R$ on account of the periodicity of the expression in (\ref{Ketto}).
 
To summarize, we have shown that every element 
$F\in {\tt H}$ must satisfy the three 
conditions\footnote{These 
conditions do not depend on the 2-form $\rho$, a reason for this 
is described at the end of sec.~3.5.} 
expressed by  (\ref{Egy}), 
the periodicity of $Y_J(x)$ in (\ref{Ketto}),  and (\ref{Harom}).
Conversely, it turns out that these conditions characterize 
${\tt H}$.
In fact, if these conditions are satisfied then  
the solution of (\ref{Elso}), (\ref{Masodik}) for $\eta$ is given by 
\be
\eta(x)=g^{-1}(x)A^F(x)g(x)-{1\over2}a^F+r(M)\big(a^F\big)\,,
\label{etasol}
\ee 
where 
\be 
r(M)\big(a^F\big)= T_\alpha r^{\alpha\beta}(M) a^F_\beta 
\label{RM}\ee 
and the matrix $r^{\alpha\beta}$ {\em is defined as the solution 
of the linear equation} 
\be
r^{\alpha\beta} +\left( W^{\phantom{\gamma}\alpha}_\gamma
-2q_\gamma^{\phantom{\alpha}\alpha}
-2q^{\alpha\theta} 
W_{\gamma\theta}\right) r^{\gamma \beta}
={1\over 2} I^{\alpha\beta} -{1\over 2} 
W^{\beta\alpha}
+q^{\alpha\beta}
- q_\gamma^{\phantom{\alpha}\alpha} 
W^{\beta\gamma}.
\label{requ}\ee
This formula of $\eta(x)=g^{-1}(x) Y^F(g(x))$ is one of the 
main results in this paper.

Some remarks are here in order. 
First, in (\ref{requ})  we suppressed the $M$-dependence of the 
various matrices  
like 
$q_\gamma^{\phantom{\alpha}\alpha}(M)=q^{\beta\alpha}(M)I_{\beta\gamma}$
and 
$W_\gamma^{\phantom{\alpha}\alpha} = 
\tr(T_\gamma M T^\alpha M^{-1})$.   
Second, in terms of the notations given at the beginning of the section, 
$r(M)$ is the linear operator on $\G$ associated with the matrix 
$ r^{\alpha \beta}(M)$.
By introducing now  the operators  $r_\pm(M)$ and $q_\pm(M)$
that correspond to the matrices 
\be
r_\pm^{\alpha\beta}(M)=r^{\alpha\beta}(M) \pm {1\over 2} I^{\alpha\beta}
\quad\hbox{and}\quad
q_\pm^{\alpha\beta}(M)=q^{\alpha\beta}(M) \pm {1\over 2} I^{\alpha\beta},
\label{rqpm}\ee 
eq.~(\ref{requ}) can be rewritten, in fact, in the 
following equivalent form: 
\be
q_+(M)\circ r_-(M)= q_-(M)\circ \Ad (M^{-1}) \circ r_+(M).
\label{requfact}\ee
The solution can be formally written as 
\be
r(M)= {1\over 2} 
\left( q_+(M) -  q_-(M)\circ \Ad (M^{-1})\right)^{-1} 
\circ 
\left(q_+(M)+ q_-(M) \circ \Ad (M^{-1})  \right).
\label{rexplicit}\ee
This shows that one must {\em define} the domain $\check G$ 
in such a way that the inverse operator above exists, which
is always possible since it becomes the identity operator at 
$M=e\in G$.
Then it is easy to see that (\ref{rexplicit}) yields 
a smooth, {\em antisymmetric} matrix function $r^{\alpha\beta}(M)$ 
on $\check G$.
By choosing $\check G\subset G$ appropriately, hence we may indeed associate
with the smooth 2-form 
$\rho$ on $\check G$ a unique, smooth map $\check G\ni M\mapsto 
r(M)\in {\mathrm End}(\G)$.
Third, we will see in the next section that the object 
\be
\hat r(M)= r^{\alpha\beta}(M) T_\alpha \otimes T_\beta \in \G\wedge  \G
\label{hatRM}\ee
 appears in 
the classical exchange algebra that encodes the Poisson brackets 
corresponding to
the symplectic form $\kappa \Omega_{chir}^\rho$ 
and it satisfies a dynamical generalization 
of the modified classical Yang-Baxter equation.
Incidentally, $\kappa \Omega_{chir}^\rho$ is {\em symplectic} 
(i.e.~nondegenerate) 
in the sense that it permits to unambiguously determine the map
${\tt H}\ni F\mapsto Y^F$, as we just saw, and ${\tt H}$ will turn out to 
contain a `complete set of functions' on $\check \M_{chir}$.

Finally, as for the derivation of eq.~(\ref{requ}), 
note that one may arrive at the special form of  the integration constant 
in formula (\ref{etasol}) of $\eta(x)$  by the expectation 
of a classical exchange algebra type PB for the field $g(x)$,
or simply by inspecting the equations that result if one writes
$\eta(x)= g^{-1}(x)A^F(x)g(x) + {\mathrm constant}$.
After introducing (\ref{etasol}) as an ansatz, 
it is not difficult to verify that (\ref{Elso}) and (\ref{Masodik}) 
reduce to (\ref{requ}). 
 
\subsection{Elements of ${\tt H}$ and their Poisson brackets}

Below we describe a large set of functions that 
are admissible Hamiltonians and apply the result
in (\ref{etasol}) to find their hamiltonian vector fields.
We shall also discuss  the interpretation 
of these hamiltonian vector fields in terms of PBs,
in particular we shall see  that the field $g(x)$ is subject 
to a quadratic exchange algebra.

Let us first study functions that depend on $g$ only through
the current $J=\kappa g' g^{-1}$.
Of course, the evaluation functions ${\cal F}_\alpha^y[g]=J_\alpha(y)$ 
do not belong to  ${\tt H}$, since $A^{{\cal F}^y_\alpha}(x)$ 
in  (\ref{Egy}) would not be a {\em smooth} function of $x$. 
Therefore we consider the  `smeared out' version 
\be 
\F_\mu:=\int_0^{2\pi}dx{\tr}\Big(\mu(x)J(x)\Big), 
\label{3.35}\ee 
where $\mu(x)$ is a $\G$-valued, smooth, $2\pi$-periodic test function. 
In this case we find that 
\be
A^{\F_\mu}(x)=\mu(x)\qquad \hbox{and}\qquad a^{\F_\mu}=0.
\ee 
The conditions expressed by (\ref{Egy}), (\ref{Ketto}) and 
(\ref{Harom}) are trivially satisfied and thus $\F_\mu\in {\tt H}$. 
The parameter $\eta (x)$ of the hamiltonian
vector field $Y^{\F_\mu}$ is $\eta (x )=g^{-1}(x )\mu (x )g(x )$,
whence 
\be
Y^{\F_\mu}\big(g(x)\big)=\mu(x)g(x),
\quad
Y^{\F_\mu}(J(x)) = [\mu(x), J(x)] + \kappa \mu'(x),
\qquad
Y^{\F_\mu}(M)=0.  
\label{3.37}\ee 
This shows in particular that $\F_\mu$ generates an infinitesimal action 
of the loop group on the phase space with respect to which $g(x)$ is an 
affine KM primary field, and the KM current $J(x)$ 
transforms according to the coadjoint action of the  
(centrally extended) loop group, as expected.
Naturally, the {\em local} functionals of $J$ defined as the integral 
over $[0, 2\pi]$  of any differential polynomial in the components of $J$,
with periodic, smooth test function coefficients, also belong to ${\tt H}$;
the corresponding hamiltonian vector fields are easy to determine.

Now we study some {\em nonlocal} functionals of the current.
Let $\E\in G$ denote the path ordered exponential integral of $J(x)$
over $[0,2\pi]$.
More precisely, we put $\E :=E(2\pi)$, where  $E(x)\in G$
is defined as the solution of  
\be\label{pathO}
\kappa E'(x)=J(x)E(x)
\quad\hbox{with}\quad 
E(0):=e\in G.
\ee 
Let $\varphi$ be an arbitrary  smooth function on $G$.
Introduce a corresponding function, $\Phi$, on the
phase space by 
\be
\Phi [g]:= \varphi(\E).
\ee 
  {}From the well-known formula of the variation of $E(x)$,
we obtain that 
\be
A^{\Phi}(x)= {1\over \kappa} \left( E(x) T^\alpha E^{-1}(x)\right) 
({\cal R}_\alpha \varphi)(\E)
\qquad
\hbox{and}\qquad
a^{\Phi}=0,
\label{AP}\ee
where ${\cal R}_\alpha$ is defined in (\ref{derLR}).
It follows  that the conditions imposed by (\ref{Egy}) and (\ref{Ketto})
are satisfied, actually from (\ref{Ketto}) we get $Y^{\Phi}(J(x))=0$.
However, $\Phi$ does {\em not} belong to ${\tt H}$ in general.  
By means of (\ref{AP}) we get that 
\be
A_\alpha^{\Phi}(0)- A^{\Phi}_\alpha(2\pi)= {1\over \kappa}
\left({\cal D}^-_\alpha \varphi\right)(\E)
\qquad ({\cal D}_\alpha^-= {\cal R}_\alpha - {\cal L}_\alpha).
\ee
Because of the condition (\ref{Harom}), 
this means that $\Phi\in {\tt H}$ precisely if $\varphi$ is an 
{\em invariant} function
on $G$ with respect to the adjoint action of $\G$ on $G$.
Examples of invariant functions are furnished by the trace
of $\E^k$ $(k=1,2,\ldots)$ in some representation.
That only the invariant functions of $\E$ are admissible 
is a well-known result in the context of current algebras,
where they provide the Casimir functions of $J$.
In our context, we obtain from  the above
that for an invariant function $\varphi$
\be
Y^{\Phi}(g(x))= {1\over \kappa} g(x) T^\alpha ({\cal R}_\alpha \varphi)(M), 
\quad
Y^{\Phi}(J(x)) =0,
\quad
Y^{\Phi}(M)=0. 
\ee 
To derive these, we used that, since $g$ and $E$ satisfy the same
differential equation,  $g(x)= E(x) g(0)$.
Hence $M= g^{-1}(0)\E g(0)$,
and for an invariant function 
\be
g^{-1}(0) T^\alpha g(0) \left({\cal R}_\alpha \varphi\right)(\E)
=T^\alpha \left( {\cal R}_\alpha \varphi\right)(M).
\ee

The monodromy matrix $M$ is not a function of $J$ alone,
but its invariant functions coincide with those of $\E$,
and we have just seen that these functions belong to ${\tt H}$.
Let us now take an arbitrary smooth function, $\psi$, on $G$ and
associate with it a function, $\Psi$,  on the chiral WZNW phase space by 
$\Psi[g]:= \psi(M)$.
Using (\ref{XM}) and the definition of
the parameter of a vector field, eq.~(\ref{XipJ}), one gets that
\be
A^{\Psi}(x)={1\over \kappa}\left( g(x) T^\alpha g^{-1}(x)\right) 
({\cal R}_\alpha \psi)(M)
\quad\hbox{and}\quad
a^\Psi={1\over \kappa} T^\alpha 
( {\cal D}^-_\alpha \psi )(M).
\ee
It follows that $\Psi \in {\tt H}$.  
For the hamiltonian vector field we obtain $Y^\Psi(J(x))=0$ and 
\be
g^{-1}(x)Y^\Psi(g(x))={1\over 2\kappa} T^\alpha  
 \left( {\cal D}^+_\alpha  \psi\right)(M)
+{1\over \kappa} T_\alpha r^{\alpha\beta}(M)
\left( {\cal D}^-_\beta \psi\right)(M).
\label{YPsi}\ee
Let us elaborate this for the functions defined by 
the matrix elements of $M$ in some representation $\Lambda$ of $G$.
We denote these matrix elements
as $M^\Lambda_{ij}$ and denote by $g^\Lambda_{ij}(x)$ the corresponding
matrix element of $g(x)$.
Now we shall use $\hat r(M)$ in (\ref{hatRM}) and
\be
\hat r_\pm(M)= \hat r(M)\pm {1\over 2} \hat I,
\quad
M_1= M\otimes 1,\quad M_2= 1\otimes M.
\ee
Then (\ref{YPsi}) can be rewritten in the tensorial form
\be
Y^{M^\Lambda_{kl}}\bigl( g^\Lambda_{ij}(x)\bigr)
={1\over \kappa} \bigl( g(x)\otimes M \, 
\hat \Theta(M)\bigr)^\Lambda_{ik, jl}
\label{gMPB}\ee
where
\be
\hat \Theta(M):= \hat r_+(M)  -M_2^{-1} \hat r_-(M) M_2
\label{Theta}\ee
is taken in the corresponding representation of $\G$, and
our notation is 
$(K\otimes L)_{ik,jl}= K_{ij} L_{kl}$. 
Furthermore, we obtain
\be
Y^{M^\Lambda_{kl}}(M^\Lambda_{ij}) = {1\over \kappa} \Big( M\otimes M\,
\hat \Delta(M)  \Big)^\Lambda_{ik,jl}
\label{MMPB}\ee
with
\be
\hat \Delta(M):=\hat \Theta(M) - M_1^{-1}\hat \Theta(M) M_1.
\label{Delta}\ee 
We shall comment on the interpretation of these equations later on.

The PB of two smooth functions $F_1$ and $F_2$ 
on a finite dimensional smooth symplectic manifold is defined by 
the standard formula
\be
\{ F_1, F_2\} = Y^{F_2}(F_1)=-Y^{F_1}(F_2)=\Omega(Y^{F_2},Y^{F_1}),
\ee
where $Y^{F_i}$ is the hamiltonian vector field associated with $F_i$
by the symplectic form $\Omega$.
The so obtained Poisson algebra is closed under pointwise
multiplication of the functions as well as under the PB.
One may formally apply the same formula in the infinite dimensional
case to the admissible smooth functions that possess a hamiltonian 
vector field.
However, it then may be a very nontrivial problem to fully specify the
set of functions that form a closed Poisson algebra, and are a
complete set in the sense that they separate the points of the phase space.
In our case, it is clear from the foregoing formulae that the 
products of 
{\em local functionals 
of the current} $J$ and the {\em smooth functions of the monodromy matrix}
$M$ form two subsets of ${\tt H}$ that are separately closed under the
PB.
Moreover, these two subsets commute 
with each other under the PB
(they  should clearly be each others centralizer in an appropriate 
Poisson algebra).
But they do not form a complete set of functions on our phase space, 
since the fundamental field $g(x)$ cannot be completely reconstructed
out of $J(x)$ and $M$.

Let us again consider a representation $\Lambda: G\rightarrow GL(V)$ of $G$. 
Since the evaluation functions $F^x_{ij}[g]= g^\Lambda_{ij}(x)$ are 
not elements of ${\tt H}$, we smear out the local field and define 
\be 
F_\phi[g]:=\int_0^{2\pi}dx {\tr}\Big(\phi(x)g^\Lambda(x)\Big), 
\label{Fphi}\ee
where $\phi: \R \mapsto {\mathrm End}(V)$ is a smooth test function.
It is then easy to see from (\ref{Egy}) that 
\bea
&&A^{F_\phi}(x) ={1\over\kappa} g(x) 
T^\alpha g^{-1}(x) 
\int_x^{2\pi}d y 
\tr\left( \phi(y) g^\Lambda(y)T^\Lambda_\alpha \right)
\quad\hbox{for}\quad x\in [0,2\pi],
\nonumber\\ 
&& a^{F_\phi}= {1\over \kappa} T^\alpha \int_0^{2\pi} dy  
\tr\left( \phi(y) g^\Lambda(y)T^\Lambda_\alpha \right).
\eea
By inspecting the condition that $Y^{F_\phi}(J(x))$ in (\ref{Ketto})
 must be periodic,
we find that $F_\phi\in {\tt H}$ for those  $\phi$ that satisfy
\be
\phi^{(k)}(0)=\phi^{(k)}(2\pi)=0\,,\qquad\qquad k=0,1,...  
\label{restrict}\ee 
Assuming that this holds, 
the hamiltonian vector field $Y^{F_\phi}$ is obtained from 
(\ref{etasol}) as 
\be
g^{-1}(x) Y^{F_\phi}(g(x))= 
{1\over\kappa} T^\alpha  
\int_x^{2\pi}d y \tr\left( T^\Lambda_\alpha \phi(y) g^\Lambda(y)\right)
 -{1\over 2} a^{F_\phi} + r(M)(a^{F_\phi}),
\quad  x\in [0,2\pi].
\label{Yg}\ee
This permits the following interpretation.
Let us define the `Poisson bracket' of the evaluation functions 
by the equality 
\be
Y^{F_\phi}( F_\chi) := \left\{ F_\chi, F_\phi\right\}:=
\int_0^{2\pi} \int_0^{2\pi} dx dy  \tr_{12}\left( \chi(x)\otimes \phi(y) 
\left\{ g^\Lambda(x) \stackrel{\otimes}{,} g^\Lambda(y)\right\}\right),
\label{localPB}\ee
where $\tr_{12}$ means the normalized trace over $V\otimes V$ and 
\be
\left\{ g^\Lambda(x)\stackrel{\otimes}{,} g^\Lambda(y)\right\}_{ik,jl}= 
\left\{ g^\Lambda_{ij}(x), g^\Lambda_{kl}(y)\right\}.
\ee  
With these definitions, formula (\ref{Yg}) of the hamiltonian vector field 
is equivalent to 
\be 
\left\{g^\Lambda(x)\stackrel{\otimes}{,} g^\Lambda(y)\right\}
={1\over\kappa}\Big(g^\Lambda(x)\otimes
g^\Lambda(y)\Big)\Big(
\hat r(M) + {1\over 2} {\hat I} \,{\mathrm sign}\,(y-x)
\Big)^\Lambda, \quad 0< x,y<2\pi.
\label{xchPB}\ee 
Indeed,  upon integration the right hand side 
of (\ref{localPB})  equals $Y^{F_\phi}( F_\chi)$
for any functions $\phi$ and $\chi$ subject to (\ref{restrict}). 
This equation  has the form of a quadratic 
exchange algebra type PB for the field $g(x)$. 
Such a classical exchange algebra is usually regarded as a classical 
analogue 
of a quantum group symmetry in the chiral WZNW model, but observe that 
in general our r-matrix is {\em monodromy dependent}.

The admissible Hamiltonians of type ${\cal F}_\mu$, $\Psi$ and $F_\phi$ 
that we studied in the above should together generate a 
closed Poisson algebra.
Although at present we cannot 
fully characterize the set of elements that belong to this algebra,
we wish to point out that the Jacobi 
identity for three functions of type $F_\phi$,
in any Poisson algebra that contains them, 
is equivalent  to the following equation for $\hat r(M)$:
\be
\big[\hat r_{12}(M),\hat r_{23}(M)\big]
+\Theta_{\alpha\beta}(M)T^\alpha_1{\cal R}^\beta \hat r_{23}(M)
+ \hbox{cycl. perm.}=
-{1\over 4} \hat f.
\label{GCDYB}
\ee
Here 
\be\label{hatf} 
\hat f=f_{\alpha\beta\gamma}T^\alpha\otimes T^\beta\otimes T^\gamma
\ee
and the cyclic permutation is over the three tensorial factors
with  
$\hat r_{23} = r^{\alpha\beta}  (1\otimes T_\alpha\otimes T_\beta)$,
$T^\alpha_1= T^\alpha \otimes 1 \otimes 1$ and so on.
Furthermore, we use the components of  
$\hat \Theta=\Theta_{\alpha\beta}T^\alpha\otimes T^\beta$
given by (\ref{Theta}), for which 
\be
\Theta_{\alpha \beta}{\cal R}^\beta =
{1\over 2} {\cal D}^+_\alpha  
+ r_\alpha^{\phantom{\alpha} \beta} {\cal D}^-_\beta.
\ee
Eq.~(\ref{GCDYB}) can be viewed as a dynamical generalization of the 
classical modified YB equation, to which it reduces 
if the r-matrix is a monodromy independent constant.
Of course, (\ref{GCDYB}) is satisfied for any $\hat r(M)$ that arises
as a solution of (\ref{requ}), since the Jacobi identity is guaranteed
by $d\Omega^{\rho}_{chir}=0$.

For later reference, let us comment here on the analogue of eq.~(\ref{GCDYB}) 
that appears in connection with chiral WZNW Bloch waves.
The space of Bloch waves\footnote{In this context    
$\G$ is  either a complex simple Lie algebra or its normal or 
compact real form.}  
in question is defined as
\be
\M_{Bloch}:=\{ b \in C^\infty(\R, G)\,\vert\,
b(x+2\pi)= b(x) e^{ \omega},
\quad 
\omega\in \A \subset \H\,\},
\label{Bloch}\ee
where ${\cal A}$ is a certain domain in a Cartan 
subalgebra
${\cal H}$ of $\G$.
There is a natural symplectic form on this space,
which is induced by the embedding
$\M_{Bloch}\subset \M_{chir}$ and is given by $\kappa \Omega_{Bloch}$ with
\be
\Omega_{Bloch}(b)=
- {1\over 2 }\int_0^{2\pi}dx\, 
 {\tr}\left(b^{-1}d b \right) \wedge
\left(b^{-1}db \right)'
-{1\over 2 }  {\tr}\left((b^{-1} d b)(0)\wedge d \omega\right).
\label{Blochform}\ee 
It is known \cite{BDF,Chu5,BBT} (for a proof in the spirit of the present 
paper, see \cite{BFP2}) that the PBs associated with (\ref{Blochform}) 
are encoded by the following classical exchange algebra: 
\be 
\Big\{b(x)\stackrel{\otimes}{,} b(y)\Big\}
={1\over\kappa}\Big(b(x)\otimes
b(y)\Big)\Big(\hat {\cal R}(\omega) 
+{1\over 2}{\hat I} \,{\mathrm sign}\,(y-x)\Big), 
\quad 0< x,y<2\pi,
\label{BlochPB}\ee 
\be
\hat {\cal R}(\omega)= {1\over 4}\sum_{\alpha} 
{\vert \alpha \vert^2}
\coth({1\over 2}\alpha(\omega)) 
E_\alpha\otimes E_{-\alpha}.
\label{Romega}\ee
The domain $\cal A$ is chosen so that $\alpha(\omega)\notin i2\pi \Z$
for any root $\alpha$ and the root vectors $E_\alpha$ satisfy
the normalization 
$\tr(E_\alpha E_{-\alpha})= {2\over \vert \alpha\vert^2}$.
As was first pointed out in \cite{BDF}, the Jacobi identity 
of the PB in this case gives rise to the equation
\be
[\hat \cR_{12}(\omega), \hat \cR_{23}(\omega)] + 
\sum_k H_1^k  {\pa \over \pa \omega^k} 
\hat \cR_{23}(\omega) + \hbox{cycl. perm.}  =-{1\over 4}\hat f,
\label{CDYB}\ee
where $\omega_k=\tr(\omega H_k)$ with respect to a basis $H_k$ of
${\cal H}$ whose dual basis is denoted as $H^k$.   
The same classical dynamical YB equation appears in other contexts 
\cite{Gervais,Feld,Avan}, too,
and has recently received lot of attention \cite{ES,EV,Liu,Lu}.

\subsection{Constant exchange r-matrices}
 
We have seen that any symplectic structure 
$\kappa \Omega^\rho_{chir}$ on $\check\M_{chir}$ 
gives rise to a  PB of the form 
(\ref{xchPB}) governed by an `exchange r-matrix' 
$\hat r(M)$, which is a solution of eq.~(\ref{GCDYB}).
Those cases for which the  exchange r-matrix is 
$M$-independent  have already been 
discussed by Falceto-Gawedzki \cite{FG} and others.
The main point \cite{FG}  here is that one can construct an appropriate 2-form
$\rho$ out of any constant, antisymmetric solution $\hat r$ of 
the modified classical YB equation,
\be
\big[\hat r_{12},\hat r_{23}\big]+\big[\hat r_{13},\hat r_{23}\big]
+\big[\hat r_{12},\hat r_{13}\big]=-{1\over 4}\hat f,
\label{mCYBE}\ee
and then {\em the same} $\hat r$ appears in the exchange algebra determined by
$\kappa \Omega_{chir}^\rho$.
We below present the construction of \cite{FG}, showing that 
in our formalism it is easy to give a complete proof as well. 

All antisymmetric solutions of  (\ref{mCYBE}) 
are known. In fact, Belavin and Drinfeld \cite{BD} classified the solutions
in the case of a complex simple Lie algebra and 
their solutions belong also to the normal real form.
For other real forms very few solutions survive 
(see Theorem 3.3 in \cite{Rawnsley}).
There is no solution for the compact real form,
because of the negative sign of the coefficient on the right hand side.
To  explain the mechanism \cite{FG} whereby 
constant exchange r-matrices appear  in the chiral WZNW model,
we first need to recall a few standard results on  eq.~(\ref{mCYBE}), 
which can be found in the reviews (e.g.~\cite{FG,BB}).

In association with a solution of (\ref{mCYBE}),
$\hat r=r^{\alpha\beta} T_\alpha \otimes T_\beta \in \G\wedge \G$,
one has the constant linear operators
$r$ and $r_\pm= r \pm {1\over 2} I$.
It follows from (\ref{mCYBE}) that the formula 
\be
[A,B]_r=[r(A),B]+[A,r(B)],
\qquad\qquad  A,B\in \G,  
\label{Rbracket}
\ee 
defines a new Lie bracket on the linear space $\G$;
the Lie algebra $(\G, [\ ,\ ]_r)$ is denoted as $\G_r$.
Then $r_\pm: \G_r \rightarrow \G$ are Lie algebra homomorphisms,
\be
[r_\pm(A),r_\pm(B)]=r_\pm\left([A,B]_r\right).
\label{homo}
\ee
Any  $A\in \G$ can be decomposed as    
\be 
A=A_+-A_-\qquad\hbox{with}\qquad A_\pm:=r_\pm(A).
\label{Xdecomp}
\ee 
As a consequence of (\ref{homo}), one has
the following equality of linear operators on $\G$, 
\be
\exp\left(\ad A_\pm\right)\circ r_\pm = r_\pm \circ 
\exp\left( \adR A\right)\qquad \forall A\in \G.
\label{intertwine}\ee
Here $\ad A$ and $\adR A$ are 
defined by $(\ad A)(B)=[A,B]$ and
$(\adR A)(B)=[A,B]_r$ for any $A,B\in \G$.
Note also that  
there exists a neighbourhood of the unit element in $G$, 
now denoted as $\check G\subset G$,
whose elements, $M\in \check G$, admit a unique decomposition in the form  
\be
M=M_+\,M_-^{-1}
\qquad\hbox{with}\qquad 
M_\pm=e^{\Gamma_\pm},
\label{xdecomp}
\ee 
where $\Gamma$ varies in a neighbourhood of zero in $\G$.
On $\check G$ one has
\be
M^{-1} dM = M_- \left( M_+^{-1} dM_+ - M_-^{-1} dM_-\right) M_-^{-1}.
\label{dMX}\ee
Let $X$ be an arbitrary vector field on $\check G$.
By means of (\ref{homo}),
(\ref{dMX})  leads to
\be
M_\pm^{-1} dM_\pm(X)=\left( M_-^{-1} M^{-1} dM(X) M_-\right)_\pm.
\label{cruci}\ee    
On the right hand side the subscript refers to the 
decomposition (\ref{Xdecomp}).
 
According to \cite{FG}, the definition
\be 
\rho(M):=\frac{1}{2}\tr \Bigl(
M_+^{-1}dM_+\wedge M_-^{-1}dM_-\Bigr)
\label{rhoFG}\ee 
yields a 2-form on $\check G$ for which
\be
d\rho = 
{1\over 6} 
\tr\left( M^{-1} dM  \wedge M^{-1} dM  \wedge 
M^{-1}dM\right) 
\qquad\hbox{on}\qquad \check G.
\label{drho}\ee
It  is straightforward to verify  (\ref{drho}) by 
using (\ref{cruci}), (\ref{homo}) and the antisymmetry of the r-matrix,
which imply e.g.~that  
$\tr\left( A_+ [B_+,C_+]\right)= \tr\left(A_- [B_-,C_-]\right)$ 
for any $A,B,C\in \G$.  

Coming now to the main point,  
let us define $\check \M_{chir}\subset \M_{chir}$ 
to be the submanifold where the monodromy matrix 
is restricted to $\check G$.
Thanks to (\ref{drho}),  
$\rho$ in (\ref{rhoFG}) yields a symplectic form $\kappa\Omega_{chir}^\rho$
by (\ref{Orho}).
It is stated in \cite{FG} (without a proof) that the PB 
(\ref{xchPB}) on $\check \M_{chir}$  that results is in this case 
governed by the same constant r-matrix $\hat r$ out of which 
$\rho$ (\ref{rhoFG})  has  been constructed. 
Our formalism permits us to prove this important result as follows.

First, we need to rewrite the 2-form $\rho$ in (\ref{rhoFG}) in the notation
used in (3.20). With the aid of (\ref{cruci}), we obtain that 
\be
q^{\alpha\beta}(M)= W^{\alpha\gamma}(M_-) r_{\gamma \theta} 
{W}^{\theta\beta}(M_-^{-1}).
\label{qFG}\ee
where we employ the notation of (\ref{Wdef}).
Here $r_{\alpha\beta}$ refers to the solution of (\ref{mCYBE}) 
that we used to define  $\rho$,
and we have to show that this monodromy independent r-matrix
also solves the defining equation of the 
exchange r-matrix, eq.~(\ref{requfact}).

As a consequence of (\ref{qFG}), for the operators $r_\pm$ and $q_\pm$ 
that appear in (\ref{requfact}) we have 
\be
q_\pm(M) = \Ad (M_-) \circ r_\pm \circ \Ad (M_-^{-1}). 
\label{qFGoper}\ee  
By using this together with  $M=M_+ M_-^{-1}$ (\ref{xdecomp}) and 
$\Ad M_\pm = \exp\left( \ad\, \Gamma_\pm \right)$,
the desired identity  (\ref{requfact}) becomes equivalent to  
\be 
r_+\circ \exp\left( -\ad\, \Gamma_- \right)\circ r_-= r_-\circ 
\exp\left(-\ad\, \Gamma_+\right)\circ r_+.
\ee 
Because of (\ref{intertwine}),
the last equation is in turn equivalent to 
\be 
r_+\circ r_-\circ \exp\left(-\adR \Gamma\right)= r_-\circ r_+\circ 
\exp\left(-\adR \Gamma\right), 
\ee 
which is obviously valid since the operators $r_+$ and $r_-$ commute.
This proves that the constant r-matrix underlying 
$\rho$ in (\ref{rhoFG}) does indeed coincide with the exchange r-matrix
associated by (\ref{requfact}) with the corresponding 
symplectic form $\kappa \Omega^\rho_{chir}$.

A well-known feature of a constant exchange r-matrix is 
that it naturally admits a Poisson-Lie action 
of the group $G$ on $\M_{chir}$.
Observe that if $\hat r$ in (\ref{xchPB}) is independent of $M$,
then the Poisson structure on $\check \M_{chir}$ smoothly extends 
to a Poisson structure on the full $\M_{chir}$.
At the same time, one can give $G$ the structure of a PL
group  by the definition \cite{Skl,Drinfeld}
\be
\{ h\stackrel{\otimes}{,} h\} ={1\over \kappa} [\hat r, h\otimes h]
\qquad
h\in G.
\label{Skly}\ee
Then one obtains a natural (left) PL action of $G$ on $\M_{chir}$ 
by the map
\be
\M_{chir}\times G \ni (g, h)\mapsto g h^{-1}\in \M_{chir}.
\label{PLaction}\ee
Indeed, this is a Poisson map 
if $\M_{chir}\times G$ carries the direct product
of the exchange algebra PB on $\M_{chir}$ and the Sklyanin bracket
(\ref{Skly}) on $G$.
In the present case, the meaning of eqs.~(\ref{gMPB})-(\ref{Delta}) is 
that $\M_{chir}\ni g \mapsto M=g^{-1}(x) g(x+2\pi)$ 
provides the `non-Abelian' momentum map \cite{FG,BB} for this PL action. 
(Of course, an equivalent right PL action is obtained 
by replacing $h^{-1}$ with $h$ and using the opposite of the PB on $G$.)

The above mechanism cannot be used to define a PL symmetry 
on $\M_{chir}$ in the case of a compact Lie group $G$,
since (\ref{mCYBE}) has no solution for a compact $\G$.
This is somewhat puzzling since as a quantum field theory  
the WZNW model is usually  
considered in the compact domain, where 
various manifestations of quantum group symmetries were found 
in the literature \cite{qgroup}.
Later we shall see that 
PL symmetries can be defined on the chiral WZNW phase 
space by certain mechanisms different from the one described above,
and they work in the compact case too.

\subsection{A parametrization of $\M_{chir}$ and classical $\G$-symmetry}

We below introduce local coordinates on the 
chiral WZNW phase space consisting of a periodic $G$-valued field 
and the logarithm, $\Gamma$, of the monodromy matrix.
This will lead us to realize the existence of a special
choice of $\rho$ such that with respect to 
$\kappa \Omega_{chir}^\rho$ 
$\Gamma$ generates an 
infinitesimal symplectic action of 
$\G$ on $\check \M_{chir}$, i.e., a classical $\G$-symmetry.
The parametrization will also shed a new light on the
notion of admissible Hamiltonians.   

If the monodromy matrix $M$ is near to $e\in G$, 
then the chiral WZNW field can be uniquely parametrized as
\be
g(x) = h(x) e^{x\Gamma},
\label{expara}\ee
where $h(x)$ is a $G$-valued, smooth, $2\pi$-periodic function
and $\Gamma$ varies in a neighbourhood of zero in $\G$,
$\check \G\subset \G$, for which the map 
$\check \G \ni\Gamma \mapsto M=e^{2\pi \Gamma} \in \check G\subset G$ 
is a diffeomorphism.
We may identify a domain in $\M_{chir}$
with the corresponding space of parameters, 
\be
\check \M_{chir}= \widetilde{G}\times \check\G=\{ (h,\Gamma)\}.
\label{expid}\ee
An easy computation gives the following formula for 
$\Omega_{chir}$ (\ref{**}) in this parametrization:
\be
\Omega_{chir} (h,\Gamma)= 
\Omega^0_{chir} (h, \Gamma) -\rho_0(\Gamma), 
\label{Omegapar}\ee
where
 \be
\Omega^0_{chir} (h,\Gamma )=
-{1\over 2}\int_0^{2\pi}dx\, \tr\left(h^{-1} dh 
\wedge (h^{-1} dh )'\right)
+d \int_0^{2\pi}dx \tr\left(\Gamma h^{-1} dh\right) ,
\label{Omega0}\ee
\be
\rho_0(\Gamma) = -{1\over 2}\int_0^{2\pi}dx\, 
\tr\left(d \Gamma \wedge de^{x \Gamma} e^{-x \Gamma}\right).
\label{rho0}\ee
Taking into account that $M=e^{2\pi \Gamma}$,
it is not difficult to verify that
\be
 d\rho_0(\Gamma) = {1\over 6} \tr
\left( M^{-1}dM\wedge M^{-1}dM\wedge M^{-1}dM\right).
\label{drho0}\ee
Upon comparison with (\ref{dOchir}),
this implies that $d\Omega^0_{chir}=0$, which one can check directly as well. 
Recalling eq.~(\ref{dOrho}), we then notice 
that the 2-form $\rho $ in (\ref{Orho})  
in this case can be 
parametrized by an  arbitrary closed 2-form
$\beta$ on $\check \G$ as 
\be
\rho(\Gamma)=\rho_0(\Gamma) + \beta(\Gamma),\qquad   
d\beta(\Gamma)=0. \quad 
\label{beta}\ee
By (\ref{Orho}) we thus have 
$\Omega^\rho_{chir}= \Omega_{chir}^0 + \beta$, 
in particular  $\Omega^{\rho_0}_{chir}=\Omega^0_{chir}$. 
In order to determine the exchange r-matrix, $\hat r_0$,  
corresponding to $\rho_0$,
we note that the integral defining $\rho_0$ can be computed in
closed form. In fact, the linear operator, $q_0$,  associated 
with its matrix in (\ref{qpara}) according to (\ref{rule}) turns out
to be\footnote{The expressions in eqs.~(\ref{qexpl}), 
(\ref{ralt}) 
are defined by the power series expansions of the corresponding complex
analytic functions around zero. For instance \cite{GR},
$2r_0= \sum_{k=1}^\infty { 2^{2k}  B_{2k} \over (2k)!} 
({1\over 2} {\cal Y})^{2k-1}$.} 
\be
q_0=\frac{2{\cal Y}+e^{-\cal Y}-e^{\cal Y}}
{2(e^{\cal Y}-1)(1-e^{-\cal Y})}
\qquad \hbox{with}\qquad
{\cal Y}:=2\pi ({\mathrm ad}\,\Gamma ).
\label{qexpl}\ee
Then from eq.~(\ref{requfact}) we find the linear 
operator
version, $r_0$,  of the
exchange r-matrix as
\be
r_0=\frac{1}{2}\coth\frac{ {\cal Y}}{2}-\frac{1}{{\cal Y}}\,.
\label{ralt}\ee
By means of  (\ref{xchPB}) this r-matrix 
defines one of the possible monodromy
dependent exchange algebras for the chiral WZWN field,
and it also represents a nontrivial solution of (\ref{GCDYB}). 
In the knowledge of the r-matrix,
the PBs containing $M$ can be determined 
straightforwardly from (\ref{gMPB}), (\ref{MMPB}).
In particular, it is easy to see that the hamiltonian
vector field generated by the function  $\Gamma_\alpha=\tr(T_\alpha \Gamma)$ 
gives rise to the  PBs  
 \be
\{ g(x), \tilde \Gamma_\alpha\} =- g(x) T_\alpha
\quad\hbox{and}\quad 
\{ \tilde \Gamma_\alpha, \tilde \Gamma_\beta\} = 
f_{\alpha\beta}^{\gamma} \tilde \Gamma_\gamma
\quad
\hbox{for}\quad 
\tilde \Gamma_\alpha:=- 2\pi \kappa \Gamma_\alpha\,.  
\label{barGamma}
\ee 
This means that  in the case of the symplectic form 
$\kappa\Omega_{chir}^0$  the logarithm of $M$ generates a  {\em classical}
${\cal G}$-symmetry  on $\check {\cal M}_{chir}$. 
Indeed, the momentum map corresponding to this symmetry
is just $\tilde \Gamma$.
A classical $\G$-symmetry 
is sometimes called
`Abelian' to contrast it with a proper (`non-Abelian') PL symmetry,
for which the symmetry group itself is endowed with a nonzero PB \cite{BB}.

The above construction is valid for any simple Lie group.
Perhaps even more surprisingly than the possibility 
to define a classical
$\G$-symmetry on $\check \M_{chir}$ for any $\G$,
in sec.~5 it will turn out
that the symplectic structure on $\check \M_{chir}$ can be
chosen so as to be compatible with any prescribed 
PL structure on $G$.

Now we explain how the 
characterization of the admissible Hamiltonians found in sec.~3.2
appears in the coordinates $(h,\Gamma)$.
For this,  we first remark that 
on account of (\ref{expid})
it is natural to represent a vector field $X$ on $\check \M_{chir}$ as
\be
X=(X_h, X_\Gamma)
\qquad X_h \in T_h \wt{G}
\qquad
X_\Gamma \in \G
\ee
with $h^{-1} X_h \in \wt{\G}$.
By regarding $h$ and $\Gamma$ as evaluation functions,
we may write $X_h = X(h)$ and $X_\Gamma=X(\Gamma)$.
Of course, the derivative $X(F)$ of function $F$ on $\check \M_{chir}$  
with respect to $X$ is still defined by means of a curve 
to which $X$ is tangent. 
Let us call a function $F$  {\em periodically differentiable}
if its derivative with respect to any $X$ exists and has the form
\be
X(F) = \langle dF,X \rangle=\tr( d_\Gamma F X_\Gamma)+
\int_0^{2\pi} \tr\left( (h^{-1} d_h F) (h^{-1} X_h)\right)
\label{XFper}\ee
with the exterior derivative 
\be
dF=(d_h F, d_\Gamma F)
\qquad
d_h F\in T_h^* \widetilde{G}
\qquad
d_\Gamma F\in \G,
\ee
where $h^{-1} d_h F\in \widetilde{\G}$ by the natural identifications.
Our point then is that the periodically differentiable functions 
coincide precisely with the admissible Hamiltonians.
To prove this,
recall that the definition of an admissible Hamiltonian 
was that its derivative exists with respect to any vector field 
and has the form (\ref{Egy}), where $A^F$ assigns to any 
$g\in \check \M_{chir}$
a smooth $\G$-valued function on  $\R$ 
such that (\ref{Ketto}) 
gives a $2\pi$-periodic function, and $a^F$ is given by (\ref{Harom}).
If we now suppose that the derivative of a function $F$ 
has the form in (\ref{Egy}), where $A^F(x)$ is a smooth function on
$\R$, 
then by inserting the parametrization (\ref{expara})
and performing a few partial integrations we obtain 
\bea
&& X(F)= \int_0^{2\pi} dx\,\tr\left( X( \kappa g' g^{-1})A^F\right) + 
\kappa \tr\left( g^{-1}(0) X(g(0)) a^F\right)\nonumber \\
&&\qquad = 
\int_0^{2\pi}dx\, \tr\left( \left(X(h) h^{-1}\right)
\left( - \kappa (A^F)' - [A^F, \kappa g' g^{-1}] \right)
 \right)\nonumber \\
&&\qquad
+\kappa \tr\left( X(\Gamma) \int_0^{2\pi}dx\, (g^{-1} A^F g)\right)
+ \kappa \int_0^{2\pi}dx\, \tr\left( [ e^{-x\Gamma} X(e^{x\Gamma}), \Gamma] 
(g^{-1}(x) A^F(x) g(x))\right) \nonumber\\
&&\qquad + 
\kappa \tr\left( \left( g^{-1}(0) X g(0)\right)
\left( a^F + g^{-1}(0)[ A^F(2\pi) - A^F(0)] g(0)\right) \right)
\label{C10}\eea
for any vector field $X$ on $\check \M_{chir}$.
Clearly, this formula can be rewritten in the form
of (\ref{XFper}) if and only if  
\be
a^F+g^{-1}(0)[ A^F(2\pi) - A^F(0)] g(0)=0,
\label{C13}\ee 
\be
- (d_h F) h^{-1}=  \kappa (A^F)' + [ A^F, \kappa g' g^{-1}]
\label{C12}\ee
and
\bea
\tr( X(\Gamma) d_\Gamma F) &=&
 \kappa \int_0^{2\pi}dx\,\tr\left( \left( X(\Gamma) +
[ e^{-x\Gamma} X(e^{x\Gamma}), \Gamma] \right)
 (g^{-1} A^F g)\right)\nonumber\\
&=& \kappa \int_0^{2\pi}dx\, \tr\left( \left( e^{-x\Gamma} 
X(e^{x\Gamma})\right)' 
(g^{-1} A^F g)(x)\right)\nonumber\\
&=&\kappa\tr\left( X(\Gamma) \int_0^{2\pi}dx\, (h^{-1} A^F h)(x) \right).
\label{C14}\eea
The last equality follows by an easy calculation, and it
implies that 
\be
d_\Gamma F= \kappa \int_0^{2\pi} \,(h^{-1} A^F h)(x).
\label{C15}\ee
We conclude that a function $F$ for which $X(F)$ has the form
of the first line in (\ref{C10}) is periodically differentiable
if and only if (\ref{C13}) is satisfied  and the right hand side of 
(\ref{C12}) defines a smooth, $2\pi$-periodic function.
In particular, all admissible Hamiltonians  are periodically 
differentiable.
Conversely, every periodically differentiable
function is admissible in the sense of sec.~3.2
since  it is possible to uniquely determine 
$(A^F, a^F)$ in terms of $(d_h F, d_\Gamma F)$
in such a way (\ref{XFper}) is converted into the first line of
(\ref{C10}). 
To achieve this, if $(d_h F, d_\Gamma F)$ are given,
one has to solve 
the differential equation (\ref{C12}) together with
the condition in (\ref{C15})  
for $A^F(x)$, and then one may define $a^F$ by (\ref{C13}).
It is not difficult to show that, since $\Gamma$ is restricted 
to $\check \G$, (\ref{C15})  
has a unique solution for the initial value 
$A^F(0)$ of $A^F(x)$, which completes the proof.

\section{Kinematical derivation of the Poisson brackets}
\setcounter{equation}{0}

In this section we rederive the PBs of the chirally
extended WZNW model using purely \lq kinematical' considerations.
Instead of explicitly inverting the symplectic form,
we postulate the natural properties of the chiral extension
(which we have established in the symplectic formalism)
and this way we can reproduce the quadratic exchange algebras 
(\ref{xchPB}) and (\ref{BlochPB}) almost effortlessly. This is 
especially so in the case of Bloch waves, where the dynamical 
r-matrix (\ref{Romega}) is determined algebraically.
The subsequent considerations are complementary to the
symplectic approach presented 
in  sec.~3 and, for Bloch waves, in \cite{BFP2}. 
The kinematical derivation sheds a new light on the origin of  
the chiral exchange algebras, and some issues are also easier 
to discuss in this approach. 

We have seen in sec.~2 that it is very natural to extend the WZNW
phase space as
\be
{\cal M}\approx{\cal M}^{sol}\quad\rightarrow\quad
{\cal M}^{ext} = \check {\cal M}_L\,\times\,\check {\cal M}_R\,,
\label{extend}
\ee
where $\check {\cal M}_L$ and 
$\check {\cal M}_R$ are two identical copies of
the chiral phase space characterized by the smooth, 
quasiperiodic chiral fields
$g_C(x+2\pi)=g_C(x)\,M_C\,,\ C=L,R$.
Since the separation of the chiral degrees of freedom is an essential
feature of the WZNW model, we shall assume that
$\check {\cal M}_L$ and $\check {\cal M}_R$ are independent and they are
equipped with the same symplectic structure (up to an  
overall sign difference, see eq. (\ref{signdif})).
The corresponding Poisson algebra  
will be supposed to contain the important Hamiltonians
${\cal F}_\mu$, $\Psi$ and $F_\phi$ studied in sec.~3.3.
The further main  assumptions  of the kinematical approach 
are that the constraints
$M_L=M_R$
are first class and the corresponding gauge 
transformations\footnote{ 
In the rest of the paper, when considering gauge transformations
or $G$-symmetries, we shall always assume
that the domain of allowed monodromy matrices, $\check G\subset G$, 
consists of full conjugacy classes in $G$. 
If this did not hold, 
everything would still be true for {\em infinitesimal} 
gauge transformations or $\G$-symmetries.}
operate according to  (\ref{bundle}) so that the WZNW
solution
$g(\sigma,\tau)=g_L(x_L)\,g_R^{-1}(x_R)$
is gauge invariant. 
These 
assumptions, 
together with
simple properties of the original  WZNW
phase space $\M^{sol}$, allow us to reproduce 
the PBs (\ref{xchPB}) and (\ref{BlochPB}).
 
\subsection{The chiral Poisson brackets for generic monodromy}
 
 {}From now on we mainly concentrate on the chiral half of the
problem and, for notational simplicity, omit the subscript $C$, 
wherever it is possible.

We start by noting that because the Fourier components of the
left KM current belong to the space of admissible
Hamiltonians, acting on the
 left  chiral field they must generate the transformation
\be
\Big\{J^n_T,g(x)\Big\}=-T\,e^{-inx}g(x)\,,
\qquad\hbox{where}\qquad
J_T^n:= \int_0^{2\pi} dx\,e^{-inx} \tr\big(T J(x)\big)\,, 
\label{primary}
\ee
which means that the chiral field is a KM primary field.
This crucial relation can be obtained by first noting that 
on the submanifold of $\M^{ext}$ defined by imposing the constraint $M_L=M_R$
(\ref{Fourier}) holds 
for the product (\ref{gsol}) that gives the WZNW solution.  
The gauge invariance
of the solution field and the fact that the different chiral pieces
completely Poisson commute then allow us to derive (\ref{primary}). 
Of course, an analogous relation is valid for the right-moving fields.

An other quantity which, by assumption, belongs to the space of
admissible Hamiltonians is the monodromy matrix $M$. From (\ref{primary}) 
it follows that
\be
\Big\{J^n_T,M\Big\}=0\,,
\label{Mcommute}
\ee
which is obvious because $M$ is invariant under the KM transformations. 
For later use we note that a quantity which Poisson commutes with
the Fourier components of the KM current must be a function of
the monodromy matrix $M$. (This is most easily seen by using the
parametrization (\ref{expara}).)

We wish to determine the PBs of the \lq smeared out' 
field\footnote{
Here some representation $\Lambda$ of $G$ is used like in (\ref{Fphi}), 
but henceforth $\Lambda$ is omitted from all notations.}
\be
F_\phi=\int^{2\pi}_0dx\tr\left(\phi(x)g(x)\right),
\label{smear}
\ee
where the matrix valued test function $\phi(x)$ satisfies (\ref{restrict}). 
Note that while in the symplectic approach of sec.~3 the fact
that $F_\phi$ is an admissible Hamiltonian follows from the
properties of the symplectic form, here it is an additional assumption. 
In order to compute the PBs with $F_\phi$ it is enough to find
\be
B_\phi(x):= \Big\{F_\phi,g(x) \Big\}
\label{Bdef}
\ee
for $0\leq x\leq2\pi$
($B_\phi(x)$ corresponds to $-Y^{F_\phi}(g(x))$ in sec.~3.3).
To constrain $B_\phi(x)$ 
we apply (\ref{primary})
to $F_\phi$ and then using the fact that $F_\phi$ is an admissible
Hamiltonian we obtain the local form 
\be
\Big\{F_\phi,J(x)\Big\}=\tr\Big(\phi(x)T^\alpha g(x)\Big)T_\alpha\,.
\label{FJ}
\ee
Comparing (\ref{Bdef}) and (\ref{FJ}) leads to the differential
equation
\be
\kappa B_\phi^\prime(x)-J(x)B_\phi(x)=\tr\Big(\phi(x)
T^\alpha g(x)\Big)T_\alpha g(x)\,,
\label{Beq}
\ee
whose solution is
\be
B_\phi(x)={1\over2\kappa}\int^{2\pi}_0dy\,{\mathrm sign}(x-y)\tr\Big(\phi(y)
g(y)T^\alpha\Big)g(x)T_\alpha+g(x)U_\phi\,,
\label{Bsol}
\ee
where
\be
U_\phi=U^\alpha_\phi T_\alpha
\ee
is a constant element of the Lie algebra. If we consider $F_{\tilde\phi}$,  
an other Hamiltonian of type $F_\phi$ belonging to a smearing function
$\tilde\phi(x)$, and use the antisymmetry of the PB
$\{F_\phi,F_{\tilde\phi}\}$, we obtain
\be
U^\alpha_\phi\tilde\phi_\alpha=-U^\alpha_{\tilde\phi}\phi_\alpha\,,
\label{anti}
\ee
where 
\be
\phi_\alpha :=\int^{2\pi}_0dx\tr\Big(\phi(x)g(x)T_\alpha\Big)
\ee
and $\tilde\phi_\alpha$ is defined analogously.
Eq.~(\ref{anti}) implies that $U^\alpha_\phi$ is a linear combination of the
integrals $\phi_\alpha$ of the form
\be
U^\alpha_\phi=-{1\over\kappa}\,r^{\alpha\beta}\phi_\beta\,,
\label{U}
\ee
where $r^{\alpha\beta}=-r^{\beta\alpha}$. 
By means of (\ref{U}), (\ref{Bsol}) becomes equivalent
to the classical exchange algebra 
\be
\Big\{g(x)\stackrel{\otimes}{,} g(y)\Big\}=
{1\over\kappa}g(x)\otimes g(y)\Big(\hat r 
+{1\over2}\hat I\,{\mathrm sign}(y-x) \Big),
\qquad 0<x,y<2 \pi\,,
\label{XXchange}
\ee
where the r-matrix 
$\hat r=r^{\alpha\beta}T_\alpha\otimes T_\beta$ is an $x$-independent
constant. Of course, as before, (\ref{XXchange}) has to be interpreted
in the distributional sense.

Although $x$-independent, the r-matrix can still depend on the 
phase space. This latter dependence can be restricted by Poisson 
commuting $J^n_T$ with (\ref{XXchange}) and applying 
the Jacobi identity. In this way we get
\be
\Big\{J^n_T,r^{\alpha\beta}\Big\}=0\,,
\ee
which implies, as explained earlier, that $\hat r$ 
must be a function of $M$ alone.

Next we consider the PBs of the monodromy matrix. 
Using (\ref{Bdef}), (\ref{Bsol}) and (\ref{U}) we get
\be
\Big\{F_\phi,M\Big\}=g^{-1}(x)\Big(B_\phi(2\pi)-B_\phi(0)M\Big)
={1\over \kappa}
\phi_\alpha\left({1\over 2}(MT^\alpha+T^\alpha M) 
 + r^{\alpha\beta}( MT_\beta - T_\beta M)\right)
\ee
and because $M$ is an admissible Hamiltonian this 
implies the local form
\be
\Big\{g(x)\stackrel{\otimes}{,} M\Big\}=
{1\over\kappa}g(x)\otimes M\,\hat\Theta(M)\,,
\label{Omega1}
\ee
where $\hat\Theta$ is given by (\ref{Theta}).
We obtain in a similar way that 
\be
\Big\{M\stackrel{\otimes}{,} M\Big\}=
{1\over\kappa}M\otimes M\,\hat\Delta(M)
\label{Omegabar1}
\ee
with $\hat\Delta$ defined in (\ref{Delta}).

To ascertain that our construction is self-consistent, we now show that
\be
{\cal C}=M_L-M_R\approx0
\label{Konstraint}
\ee
is a first class constraint on $\M^{ext}$
and the WZNW solution is gauge invariant.
In fact, ${\cal C}$ is first class
since
\bea
\Big\{{\cal C}\stackrel{\otimes}{,} {\cal C}\Big\}&
=&\Big\{M_L\stackrel{\otimes}{,} M_L\Big\}+\Big\{M_R\stackrel{\otimes}{,}
M_R\Big\}\nonumber\\
&=&{1\over\kappa}\big(M_L\otimes M_R\Big)
\hat\Delta(M_L)-{1\over\kappa}\Big(M_R\otimes M_R\Big)
\hat\Delta(M_R)\approx0\,.\nonumber
\eea
Similarly, 
the gauge invariance of $g(\sigma, \tau)$ in (\ref{gsol}) can 
be shown as follows:
\bea
\kappa\,\Big\{g(\sigma,\tau)\stackrel{\otimes}{,} {\cal C}\Big\}&
=&\Big(g_L(x_L)\otimes M_L\Big)
\hat\Theta(M_L)\Big(g_R^{-1}(x_R)\otimes1\Big)\nonumber\\
&-&\Big(g_L(x_L)\otimes M_R\Big)
\hat\Theta(M_R)\Big(g_R^{-1}(x_R)\otimes1\Big)\nonumber\\
&\approx&
\Big(g_L(x_L)\otimes M\Big)
\Big(\hat\Theta(M_L)-\hat\Theta(M_R)\Big)\Big(g_R^{-1}(x_R)\otimes1\Big)
\approx0\,.
\nonumber
\eea
Here the notation $\approx$ indicates `weak'
equality, i.e., equality on the constrained manifold,
and we used the assumption that $\check {\cal M}_L$ and $\check {\cal M}_R$
are independent and carry opposite PBs.

For later use we mention an additional consistency check. Consider
the path ordered integral 
${\cal E}=E(2\pi)$ defined in (\ref{pathO}).
Since ${\cal E}$ and $M$ are related by conjugation, 
their invariants coincide:
\be
\varepsilon_N =\tr\left({\cal E}^N\right)=\tr\left(M^N\right)=m_N,
\label{Ntrace}
\ee
and since the PBs of the Hamiltonians $\varepsilon_N$ can
be calculated unambiguously using the KM algebra only, the following
relation must hold:
\be
\Big\{g(x),m_N\Big\}=\Big\{g(x),\varepsilon_N\Big\}={N\over\kappa}g(x)T^\alpha
\tr\Big(M^NT_\alpha\Big)\,.
\label{mNpN}
\ee
It is easy to verify by using (\ref{Theta}) in (\ref{Omega1})
that (\ref{mNpN}) is indeed satisfied. 

To summarize, by postulating the $M_L=M_R$ constraint to be first
class, as well as the admissible Hamiltonian nature of the Fourier 
components of the current, the smeared out chiral field and the 
monodromy matrix,  
we have established that the possible extensions of the WZNW phase
space correspond to the quadratic exchange algebra
(\ref{XXchange}) with some monodromy dependent exchange r-matrix.
Of course the classical exchange algebra can only
provide a valid PB if it satisfies the Jacobi identity.
This is guaranteed, by construction, if the r-matrix is obtained as a
solution of (\ref{requfact}). In the present approach, we have to {\em impose}
the Jacobi condition and, as mentioned in sec.~3.3, this leads to 
the  dynamical YB equation (\ref{GCDYB}).
The chiral extensions of the WZNW model are thus characterized by the 
solutions of this equation.

Most known solutions of (\ref{GCDYB}) are local in that they are defined only
on a proper, open submanifold $\check G\subset G$.
This is obviously the case for the solutions obtained by solving 
(\ref{requfact}) for the r-matrix, starting from a q-matrix 
representing by (\ref{qpara}) a local solution of (\ref{dOrho}). 
On the other
hand, as the example of constant r-matrices shows, there are
global solutions as well. Since constant solutions exist 
for non-compact groups only, an interesting open question
is whether there exist globally defined exchange 
r-matrices also for compact groups. 
We hope to return to this question in a future study.
 
We end this subsection by discussing a generalization of the \lq gauge'
freedom (\ref{bundle}) of the chiral extension. It is clear that
the gauge transformed chiral fields given by
\be
g(x)\ \rightarrow\ (\P g)(x)=g(x)\,p(M)
\label{gau}
\ee
with an arbitrary {\it monodromy dependent} group element $p(M)$
correspond to the same point in the physical phase space after the
projection (\ref{theta}), provided we apply the same gauge transformation
to both chiral fields. In other words, both the original fields, $g(x)$,
and the gauge transformed fields, $\tilde g(x):=g(x)\, p(M)$, are
smooth quasiperiodic $G$-valued \lq coordinates' of the same point
in the physical phase space.

The multiplication law for two elements, $\P_1, \P_2$, of this
huge gauge group is given by
\be
p_{12}(M)=p_2(M)\, p_1(\P_2 M)\,,
\label{multi}
\ee
where $p_1,p_2$ and $p_{12}$ correspond to
$\P_1,\P_2$ and $\P_{12}:=\P_1 \P_2 :=\P_1\circ \P_2$,
respectively. Here
\be
\P M:=p^{-1}(M)Mp(M)
\label{gauM}
\ee
is how the monodromy matrix itself is transformed under a gauge 
transformation. To qualify as an element of the gauge group we must also
require that the inverse transformation exists. This is equivalent to
requiring that the group valued function, $\bar p(M)$, corresponding
to the inverse element, $\P^{-1}$, exists and solves
\be
p(M)\bar p(\P M)=e\,.
\label{inv}
\ee

In terms of the new \lq coordinates' $\tilde g(x)$ defined by 
(\ref{gau}) the exchange
algebra has the same form as (\ref{XXchange}) with a gauge
transformed exchange r-matrix, $\hat{\tilde r}$.
On account of the Leibniz rule of the PB, one finds that 
\be
\hat{\tilde r}=p^{-1}(M)\otimes p^{-1}(M)\Big(
\hat r(M)+\Theta^{\alpha\beta}\big[T_\alpha\otimes{\cal A}_\beta-
{\cal A}_\beta\otimes T_\alpha\big]+\Delta^{\alpha\beta}{\cal A}_\alpha
\otimes{\cal A}_\beta\Big)p(M)\otimes p(M)\,,
\label{rtilde}
\ee
where ${\cal A}_\alpha:=({\cal R}_\alpha p)p^{-1}$ and
$\hat{\tilde r}$ should be expressed as a function of the
gauge transformed monodromy $\tilde M:=\P M$.
Since the Jacobi identity of the exchange algebra (\ref{XXchange}) is
independent of the coordinates used,  
it is clear that the solutions of the dynamical YB equation 
(\ref{GCDYB})
are transformed into each other by the elements of
the gauge group and therefore
can be classified up to gauge transformations.

\subsection{Diagonal monodromy}

Below we briefly  outline a kinematical derivation of the PBs
on the space of chiral WZNW Bloch waves, $\M_{Bloch}$ defined in (\ref{Bloch}).
In order to emphasize their diagonality, we denote the 
monodromy matrices of the Bloch waves, $b(x)$,  here by $D$,
\be
D=e^{\omega^kH_k}\,,
\ee
where $H_k$ are the basis elements of a splitting Cartan subalgebra of $\G$.
We will also use the derivatives
$\partial_k={\partial\over\partial\omega^k}\,$.

The assumptions and the main steps of the construction (with obvious
modifications for the diagonal case) are the same as discussed above
for the general case.
Now we obtain a  classical exchange algebra of the form 
\be
\Big\{b(x)\stackrel{\otimes}{,} b(y)\Big\}=
{1\over\kappa}b(x)\otimes b(y)\Big(\hat r(\omega) 
+{1\over2}\hat I\,{\mathrm sign}(y-x)\Big),
\qquad 0<x,y<2 \pi .
\label{BXchange}
\ee
The main difference comes from requiring
(\ref{mNpN}), because, unlike in the general case where it was a consequence
of the exchange algebra, here it completely determines the PBs
of the monodromy matrix:
\be
\Big\{b(x)\stackrel{\otimes}{,} D\Big\}={1\over\kappa}\big(
b(x)\otimes D\big)\,\big(H_k\otimes H^k\big)\,.
\label{Omega1D}
\ee
On the other hand, the analogue of (\ref{Omega1}) implied by  
(\ref{BXchange}) now reads explicitly as 
\be
\Big\{b(x)\stackrel{\otimes}{,} D\Big\}={1\over\kappa}
\left(b(x) \otimes D\right) 
\Big(\hat r(\omega)+{1\over 2}\hat I -\big(1\otimes D^{-1}\big) \big(
\hat r(\omega)-{1\over 2}\hat I\big)\big(1\otimes D\big)\Big).
\label{Omegaeq}
\ee
The comparison of the last two equations fixes 
the exchange r-matrix almost completely:
\be
\hat r(\omega)=\hat {\cal R}(\omega)+\hat X(\omega)\,,
\label{rRX}
\ee
where  $\hat{\cal R}(\omega)$ is given by (\ref{Romega})
and $\hat X(\omega)$ is an antisymmetric purely Cartan piece,
\be
\hat X(\omega)=X^{kl}(\omega)H_k\otimes H_l\,,\qquad\qquad
X^{kl}(\omega)=-X^{lk}(\omega)\,.
\label{Xpiece}
\ee
Thus we have determined the exchange  r-matrix algebraically 
up to the Cartan piece.

The Jacobi identity takes the following form for the diagonal case:
\be
\big[\hat r_{12}(\omega),\hat r_{23}(\omega)\big]+
H^k_1\,\partial_k\hat r_{23}(\omega) + \hbox{cycl. perm.} =-{1\over4}\hat f.
\label{CDYBr}
\ee
This is the celebrated CDYB equation, whose neutral
solutions have been classified  in \cite{EV}.
The r-matrix (\ref{rRX}) satisfies (\ref{CDYBr}) if
\be
\partial_kX_{lm}+\partial_lX_{mk}+\partial_mX_{kl}=0.
\ee
Therefore there exists a `gauge
vector' $V_k(\omega)$ such that
\be
X_{kl}(\omega)=\partial_kV_l(\omega)-\partial_lV_k(\omega).
\label{XV}
\ee
With the help of $V_k(\omega)$ we can introduce the gauge transformed
 chiral field
\be
\tilde b(x)=b(x)e^{-V_k(\omega)H^k},
\label{gauD}
\ee
which has diagonal monodromy and satisfies the `standard'
exchange algebra
\be
\Big\{\tilde b(x)\stackrel{\otimes}{,} \tilde b(y)\Big\}={1\over\kappa}
\tilde b(x)\otimes \tilde b(y)\Big(\hat{\cal R}(\omega)
+{1\over2}\hat I\,{\mathrm sign}(y-x)\Big)\,,
\qquad 0<x,y<2\pi\,.
\label{XXchangeD}
\ee

It has been mentioned that the PB (\ref{XXchangeD}) follows \cite{BFP2} from 
the symplectic form $\kappa \Omega_{Bloch}(\tilde b)$ on $\M_{Bloch}$
given by (\ref{Blochform}).
Upon the substitution (\ref{gauD}), the symplectic form 
gets shifted by the exact 2-form $\kappa X_{kl}(\omega)\,d\omega^k\wedge
d\omega^l$ and the shifted symplectic form corresponds of course 
to the exchange algebra (\ref{BXchange}) with $\hat r(\omega)$ in (\ref{rRX}).
It is also easy to see that the family of the symplectic forms
$\kappa \left( \Omega_{Bloch}(b) + 
X_{kl}(\omega) \omega^k \wedge \omega^l\right)$
 on $\M_{Bloch}$ results as the reduction of the family of symplectic forms 
$\kappa \Omega_{chir}^\rho$ to diagonal monodromy.
 
\section{Exchange r-matrices with Poisson-Lie symmetry} 
\setcounter{equation}{0}

We studied in sec.~3.4 the Poisson-Lie  action of the group $G$
on the chiral phase space for the special case of a constant r-matrix
playing the r\^ole of the exchange r-matrix. As mentioned there,
this PL action is restricted to the case of complex or real, non-compact
groups, since there is no constant solution of (\ref{GCDYB}) for compact
groups. In this section we consider a set of more general PL actions 
which also work for the physically most interesting case of compact groups.

It is clear from the examples studied so far that the PL action we are
looking for is a kind of \lq hidden' symmetry, extending and centralizing
the Kac-Moody  symmetries in the total symmetry group of the
chiral WZNW model. More precisely, we require that
\begin{itemize}
\item the KM currents are invariant under the PL action;
\item the PL action commutes with the KM transformations.
\end{itemize}

It is not difficult to see that because of the above two requirements 
the PL action as a nonlinear group action on the chiral phase space
has to be a special case of the gauge transformations
discussed in sec.~4.1:
\be
g(x)\ \rightarrow\ 
\P_hg(x)=g(x)p(M,h)
\qquad\quad \forall h\in G.
\label{PLakcio}
\ee
The $G$-valued function $p(M,h)$ is chosen so that
the group multiplication law $\P_h\P_k=\P_{hk}$ is satisfied:
\be
p(M,k)p(\P_kM,h)=p(M,hk)\,,
\ee
where the induced action on the monodromy matrix is
$\P_kM=p^{-1}(M,k)Mp(M,k)$. One must also require
that $p(M,e)=e$.

The simplest case is the \lq standard' (left) action 
\be
\S_hg(x)=g(x)h^{-1}
\label{standard}
\ee
corresponding to $p(M,h)=h^{-1}$. A family of \lq trivial'
actions is obtained by conjugating the standard action in the gauge 
group by an arbitrary element $\P$ (see (\ref{gau})):
\be
\tilde \S_h=\P^{-1}\S_h\P .
\label{trivial}
\ee
In terms of the corresponding $G$-valued functions we have
\be
\tilde s(M,h)=p(M)h^{-1}\bar p(h\cdot \P M\cdot h^{-1})\,.
\label{triviomega}
\ee
 
In practice it is useful to consider the infinitesimal version
of (\ref{PLakcio}). For the parametrization $h=e^{u^\alpha T_\alpha}$
we have
\be
\P_hg(x)=g(x)-u^\alpha X_\alpha g(x)+{\cal O}(u^2)\,,
\ee
where the infinitesimal generators are of the form
\be
X_\alpha g(x)=-\zeta^{\beta\phantom{\alpha}}_{\phantom{\alpha}\alpha}(M)
g(x)T_\beta
\label{infi}
\ee
with some monodromy dependent coefficients
$\zeta^{\beta\phantom{\alpha}}_{\phantom{\alpha}\alpha}(M)$, 
and satisfy the commutation relations $[X_\alpha,X_\beta]=
f^{\phantom{\alpha\alpha}\gamma}_{\alpha\beta\phantom{\alpha}}
X_\gamma$. This latter equation can be expressed as a requirement
on the coefficients
$\zeta^{\beta\phantom{\alpha}}_{\phantom{\alpha}\alpha}$ as
\be
\zeta^{\lambda\phantom{\alpha}}_{\phantom{\alpha}\alpha}
{\cal D}^-_\lambda
\zeta^{\omega\phantom{\alpha}}_{\phantom{\alpha}\beta}
-\zeta^{\lambda\phantom{\alpha}}_{\phantom{\alpha}\beta}
{\cal D}^-_\lambda
\zeta^{\omega\phantom{\alpha}}_{\phantom{\alpha}\alpha}
+f^{\phantom{\alpha\alpha}\omega}_{\kappa\lambda\phantom{\alpha}}
\zeta^{\kappa\phantom{\alpha}}_{\phantom{\alpha}\alpha}
\zeta^{\lambda\phantom{\alpha}}_{\phantom{\alpha}\beta}
+f^{\phantom{\alpha\alpha}\lambda}_{\alpha\beta\phantom{\alpha}}
\zeta^{\omega\phantom{\alpha}}_{\phantom{\alpha}\lambda}=0\,.
\label{phantoms}
\ee
Clearly the simplest solution of (\ref{phantoms}) is the standard
one, $\zeta^{\beta\phantom{\alpha}}_{\phantom{\alpha}\alpha}
=-\delta^{\beta\phantom{\alpha}}_{\phantom{\alpha}\alpha}$.

The infinitesimal generators  are also useful in studying the question of
trivialization. The infinitesimal form of (\ref{triviomega}) is
\be
p T_\alpha+
\zeta^{\lambda\phantom{\alpha}}_{\phantom{\alpha}\alpha} 
({\cal D}^-_\lambda+T_\lambda) p  =0\,.
\label{inftrivi}
\ee
It is easy to see that the consistency conditions of this set
of partial differential equations are precisely the equations
(\ref{phantoms}), but it is not clear if 
all possible nonlinear actions $\P_h$ can be trivialized
in the form (\ref{trivial}) or not.

The next question is which of the above nonlinear
group actions are Poisson-Lie? 
Following \cite{BB}, we recall
that a Lie group (or algebra) 
action characterized by infinitesimal generators
$X_\alpha$ is Poisson-Lie, if for any pair of phase space functions
$F_1,F_2$ the PBs satisfy the relations
\be
\Big\{X_\alpha(F_1),F_2\Big\}+\Big\{F_1,X_\alpha(F_2)\Big\}-
X_\alpha\Big(\big\{F_1,F_2\big\}\Big)=
-
\tilde f^{\beta\gamma\phantom{\alpha}}_{\phantom{\alpha\alpha}\alpha}
\, X_\beta(F_1) \, X_\gamma(F_2)\,,
\label{PLsymmetry}
\ee
where the pair of structure constants $(
f^{\phantom{\alpha\alpha}\gamma}_{\alpha\beta\phantom{\alpha}},
\tilde f^{\beta\gamma\phantom{\alpha}}_{\phantom{\alpha\alpha}\alpha})$
together define a Lie bi-algebra.
Now applying this to $F_1=g(x)$, $F_2=g(y)$ and parametrizing the
exchange r-matrix as
\be
r^{\alpha\beta}=k^{\kappa\lambda}
\zeta^{\alpha\phantom{\alpha}}_{\phantom{\alpha}\kappa}
\zeta^{\beta\phantom{\alpha}}_{\phantom{\alpha}\lambda}
\label{k-matrix}
\ee
we find that the infinitesimal action (\ref{infi}) will be PL if
\be
\zeta^{\alpha\phantom{\alpha}}_{\phantom{\alpha}\kappa}
\zeta^{\beta\phantom{\alpha}}_{\phantom{\alpha}\lambda}
\Big(
\zeta^{\sigma\phantom{\alpha}}_{\phantom{\alpha}\gamma}
{\cal D}^-_\sigma k^{\kappa\lambda}
+k^{\kappa\sigma}
f^{\phantom{\alpha\alpha}\lambda}_{\sigma\gamma\phantom{\alpha}}
+k^{\sigma\lambda}
f^{\phantom{\alpha\alpha}\kappa}_{\sigma\gamma\phantom{\alpha}} +\kappa 
\tilde f^{\kappa\lambda\phantom{\alpha}}_{\phantom{\alpha\alpha}\gamma}\Big)
={1\over2}\,\Big({\cal D}^{+\alpha}
\zeta^{\beta\phantom{\alpha}}_{\phantom{\alpha}\gamma}
-{\cal D}^{+\beta}
\zeta^{\alpha\phantom{\alpha}}_{\phantom{\alpha}\gamma}\Big)\,.
\label{BBeq1}
\ee
It is well-known that all simple PL groups are coboundary.
This means that the PL structure on $G$ is  given by
the Sklyanin bracket,
\be
\big\{h\stackrel{\otimes}{,}h\big\}={1\over\kappa}\,
\Big[\hat R,h\otimes h\Big],
\label{PBgroup}
\ee
for which the structure constants of the induced dual Lie algebra are 
\be
 \tilde f^{\beta\gamma\phantom{\alpha}}_{\phantom{\alpha\alpha}\alpha}=
{1\over \kappa} \left(R^{\sigma\beta}
f^{\phantom{\alpha\alpha}\gamma}_{\sigma\alpha\phantom{\alpha}}
+
R^{\gamma\sigma}
f^{\phantom{\alpha\alpha}\beta}_{\sigma\alpha\phantom{\alpha}}\right),
\label{cobo}
\ee
where  $\hat R=R^{\alpha\beta} T_\alpha \otimes T_\beta\in \G\wedge G$
is some  {\em constant, antisymmetric}  r-matrix.
The r-matrix that occurs here is an arbitrary 
solution of the (modified) classical YB equation,
\be
\big[\hat R_{12},\hat R_{23}\big] +
\hbox{cycl. perm.}=-\nu^2\hat f 
\label{CDYBconst}
\ee
with some constant parameter $-\nu^2$.
This parametrization of the right hand side will prove useful below.
The value $\nu=0$ is also allowed and for real Lie groups
$\nu^2$ is of course real.
 {}From the classification \cite{BD,Rawnsley}  of the solutions we recall
that for a compact simple Lie algebra $\nu$ must be purely imaginary or zero,
while for the maximally non-compact (split) real forms $\nu$ is real.  

Let us introduce $K^{\alpha\beta}$ by writing
\be
k^{\alpha\beta}(M)=K^{\alpha\beta}(M)+R^{\alpha\beta}.
\label{Kobo}
\ee
Then (\ref{BBeq1}) can be reduced to
\be
\zeta^{\alpha\phantom{\alpha}}_{\phantom{\alpha}\kappa}
\zeta^{\beta\phantom{\alpha}}_{\phantom{\alpha}\lambda}
\Big(
\zeta^{\sigma\phantom{\alpha}}_{\phantom{\alpha}\gamma}
{\cal D}^-_\sigma K^{\kappa\lambda}
+K^{\kappa\sigma}
f^{\phantom{\alpha\alpha}\lambda}_{\sigma\gamma\phantom{\alpha}}
+K^{\sigma\lambda}
f^{\phantom{\alpha\alpha}\kappa}_{\sigma\gamma\phantom{\alpha}}
\Big)
={1\over2}\,\Big({\cal D}^{+\alpha}
\zeta^{\beta\phantom{\alpha}}_{\phantom{\alpha}\gamma}
-{\cal D}^{+\beta}
\zeta^{\alpha\phantom{\alpha}}_{\phantom{\alpha}\gamma}\Big)\,.
\label{BBeq2}
\ee
This depends on how the infinitesimal generator $X_\alpha$
is parametrized in terms of 
$\zeta^{\beta\phantom{\alpha}}_{\phantom{\alpha}\alpha}$, but 
the explicit reference to the dual structure constants (\ref{cobo}) 
has disappeared.

 {}From now on we will concentrate on the simplest case corresponding
to the standard action 
$\zeta^{\beta\phantom{\alpha}}_{\phantom{\alpha}\alpha}
=-\delta^{\beta\phantom{\alpha}}_{\phantom{\alpha}\alpha}$. In this case
\be
r^{\alpha\beta}(M)=K^{\alpha\beta}(M)+R^{\alpha\beta}
\label{standardcase}
\ee
and (\ref{BBeq2}) reduces to
\be
{\cal D}^-_\alpha K=\big[K,{\T}_\alpha\big]\,,
\qquad\hbox{where}\quad{\T}_\alpha:={\mathrm ad}\,T_\alpha\,,
\label{equi1}
\ee
that is, $({\T}_\alpha)^{\phantom{\alpha}\lambda}_{
\sigma\phantom{\alpha}}=
f^{\phantom{\alpha\alpha}\lambda}_{\sigma\alpha\phantom{\alpha}}$.
Eq.~(\ref{equi1}) is the infinitesimal form of 
\be
\hat K(hMh^{-1})=(h\otimes h)\hat K(M)(h^{-1}\otimes h^{-1})\,,
\label{equi2}
\ee 
which expresses the equivariance of 
$\hat K(M)=K^{\alpha\beta}(M) T_\alpha\otimes T_\beta$ 
under the action (\ref{standard}). 
One may also verify directly that the standard left 
action of $G$ equipped with the PB (\ref{PBgroup})
is PL for the exchange r-matrix (\ref{standardcase}) if
$\hat K(M)$ is equivariant.

So far we have established that the action (\ref{standard})
of $G$ on $\check \M_{chir}$  is PL if $r(M)$ is the sum of a
constant r-matrix $R$ and an equivariant piece $K(M)$. 
Of course,
the exchange r-matrix (\ref{standardcase}) has to be a solution
of the dynamical YB equation (\ref{GCDYB}). Using
(\ref{equi1}), (\ref{GCDYB}) can be rewritten as
\be
-(K{\T}^\alpha K)^{\beta\gamma}+{1\over2}{\cal D}^{+\alpha}
K^{\beta\gamma}+\hbox{cycl.\ perm.}=\big(\nu^2-{1\over4}\big)f^{\alpha
\beta\gamma}
\label{dyneq}
\ee
in this special case, where the cyclic permutations are over
the upper indices $\alpha,\beta,\gamma$.
The search for solutions of (\ref{dyneq}) is made feasible  by the
observation that any analytic function of ${\cal Y}$ (defined
in (\ref{qexpl})) is equivariant. We show in Appendix B that a
solution of (\ref{dyneq}) is given by the analytic function
\be
K={1\over2}\coth{{\cal Y}\over2}-\nu\coth\left({\nu{\cal Y}}\right)\,.
\label{soln}
\ee
This formula is valid on a domain $\check G$ around $e\in G$ 
where the exponential parametrization is applicable and the
power series of the above expression converges.  
The derivation of the exchange r-matrix given by 
(\ref{standardcase}), (\ref{soln}), 
which is compatible with the standard action of the PL group $G$ equipped 
with the PB (\ref{PBgroup}), is one of our main results.

We conclude this section with a couple of remarks 
related to the above r-matrices. 
We first note that for $\nu=0$ (\ref{soln}) is understood as the appropriate
limit and therefore for $R=0$ we recover from (\ref{standardcase}) 
the exchange r-matrix $r_0$ (\ref{ralt}) together with the
classical $\G$-symmetry  discussed in sec.~3.5.
We can also have $\nu=0$ in correspondence 
with any antisymmetric solution $R\neq 0$ of the classical YB equation.
For $\nu=1/2$ we get $K=0$ and the dynamical  r-matrix (\ref{standardcase}) 
then reduces to one of the constant r-matrices treated  in sec.~3.4.
For compact groups all solutions (\ref{soln}) are strictly dynamical 
(non-constant), since in this case $-\nu^2 \geq 0$. 
Finally, we remark that the existence  of a suitable 
local 2-form $\rho $ corresponding to the r-matrix  
(\ref{standardcase}), (\ref{soln}) is guaranteed by the
solvability of (\ref{requfact}).

We have seen that the classical $\G$-symmetries discussed in sec.~3.5 and 
the special PL symmetries treated in sec.~3.4  are generated by momentum maps.
Without going into details, we wish to mention that 
it is possible to show the existence of  
a non-Abelian momentum map also in the general case of the above 
PL symmetries. 
The momentum map is given by a function $m(M)$ on $\check \M_{chir}$ 
(depending on the monodromy matrix only), 
which  takes its values in the dual PL group $G^*$ in such a way that for
all phase space functions
$F$
\be
X_\alpha F=\Big(m^{-1}\big\{m,F\big\}\Big)_\alpha\,,
\label{mommap1}
\ee
where the $()_\alpha$ component on the right hand side is evaluated 
in the dual Lie-algebra, ${\cal G}^*$. 
Moreover, the $G^*$-valued momentum map satisfies the PBs
\be
\{m\stackrel{\otimes}{,}m\}=\eta^*(m) m\otimes m\,,
\label{mommap2}
\ee
where the Poisson tensor 
$\eta^*(m)\in{\cal G}^*\wedge{\cal G}^*$ is chosen 
so that (\ref{mommap2}) defines the Poisson structure
of the dual PL group. 
(For an explanation of these notions, see e.g.~\cite{BB}.)

It is in principle possible to use the the momentum map 
construction also backwards.
If there is a $G^*$-valued function $m$ on the phase space
satisfying (\ref{mommap2}), then using (\ref{mommap1}) to define an
infinitesimal generator $X_\alpha$ one obtains that
\begin{itemize}
\item the infinitesimal generators represent
the Lie algebra, $[X_\alpha,X_\beta]=f_{\alpha\beta{\phantom{\alpha}}}^
{\phantom{\alpha\alpha}\gamma}X_\gamma$,
\item the Lie algebra action is PL, i.e.,  (\ref{PLsymmetry}) holds.
\end{itemize}

Let now suppose that a {\em compact} simple Lie algebra 
$\G$ acts on a phase space as a classical  symmetry 
generated by a $\G^*$-valued, equivariant (`Abelian') momentum map.
In this situation one can always define also an infinitesimal 
PL action of the group $G$ 
equipped with the so called {\em standard} PL structure.
This is a consequence of the fact \cite{weinstein} 
that there exists a diffeomorphism between ${\cal G}^*$ and $G^*$ 
that converts the natural linear Poisson structure on $\G^*$
into the standard PL structure on $G^*$.  
Applying this to the classical $\G$-symmetry studied in sec.~3.5,
we can thus find a map, 
\be
{\cal G}^*\ni\tilde \Gamma\mapsto  m(\tilde \Gamma)\in
G^*,
\label{weinmap}
\ee
where $\tilde \Gamma: \check \M_{chir}\rightarrow \G^*$ 
is given by (\ref{barGamma}), such that
$m(\tilde\Gamma)$ satisfies (\ref{mommap2}) 
with respect to the standard PL structure.
The resulting $G^*$-valued momentum map then generates a 
PL action on the phase space $\check \M_{chir}$ as outlined above. 

This somewhat surprising construction is not specific to the
chiral WZNW phase space, since it can be used whenever one has 
a classical $\G$-symmetry based on a compact simple Lie algebra.
When applying it to the chiral WZNW phase space, 
the Lie algebra action (\ref{mommap1}) constructed using the
momentum map (\ref{weinmap}) will be different from the standard 
one (\ref{standard}).
It is an interesting question whether this Lie algebra action can
be gauge transformed to the standard form and, if 
the answer is positive,  to find the
corresponding gauge transform of 
the r-matrix $r_0$ (\ref{ralt}).  
We wish to discuss this in a future publication.

\section{Interpretation in terms of Poisson-Lie groupoids}  
\setcounter{equation}{0}

The CDYB equation (\ref{CDYBr}) can be 
regarded as the guarantee of the Jacobi identity in
a PL groupoid \cite{EV}. 
Below we show that eq.~(\ref{GCDYB}) 
admits  an analogous interpretation.
For this, we introduce a family of 
PL groupoids in such a way that a subfamily 
of these is naturally associated with the possible 
PBs on the chiral WZNW phase space.
Remarkably, these groupoids are finite dimensional Poisson manifolds  
that encode practically all information about the
infinite dimensional chiral WZNW PBs.
 
Roughly speaking, a groupoid is a set, say $P$,  
endowed with a `partial multiplication'
that behaves similarly to a group multiplication in 
the cases when it can be performed.
To understand the following construction one does not 
need to know details of the notion of a groupoid
(see e.g.~\cite{Gpoid}), since we shall only use the most 
trivial example of such a structure, for which 
\be
P= S \times G \times S = \{ (M^F, g, M^I) \},
\label{Pset}\ee 
where $G$ is a group and $S$ is some set. 
The partial multiplication is defined for those triples 
$(M^F, g, M^I)$ and $(\bar M^F, \bar g, \bar M^I)$ for which 
$M^I=\bar M^F$,
and the product is 
\be
(M^F, g, M^I) (\bar M^F, \bar g, \bar M^I):= (M^F, g\bar g, \bar M^I)
\quad\hbox{for}\quad M^I=\bar M^F.
\label{Pmult}\ee 
In other words, the graph of the partial multiplication
is the subset of 
\be
P\times P\times P =
\{ (M^F, g, M^I)\} \times \{ (\bar M^F, \bar g, \bar M^I)\}
\times \{ (\hat M^F, \hat g, \hat M^I)\} 
\ee
defined by the constraints
\be
M^I=\bar M^F,
\quad
\hat M^F= M^F,
\quad
\hat M^I=\bar M^I,
\quad
\hat g= g\bar g,
\label{graph}\ee
where the hatted triple encodes the components of the product. 
A PL groupoid \cite{Wei} $P$ is a groupoid and a Poisson manifold 
in such a way that the graph of the partial
multiplication is a {\em coisotropic} submanifold of  
$P\times P\times P^-$, where $P^-$ denotes the  manifold
$P$ endowed with the opposite of the PB on  $P$.
In other words, this means that {\em the constraints that define 
the graph are first class}. 
This definition 
reduces to that of a PL group in the particular 
case for which the set $S$ in (\ref{Pset}) consists of a single point.

In the interpretation of (\ref{CDYB}) given in \cite{EV} the groupoid 
$P$ is of the form above with $S$ taken to be a domain in 
the dual of a Cartan subalgebra of a simple Lie group $G$.
By thinking about a generic monodromy matrix, 
we  now take $P$ to be
\be
P= \check G \times G \times \check G,
\ee
where $\check G$ is some open domain in $G$.
On this $P$, we postulate a PB  $\{\ ,\ \}_P$ defined, 
by using the usual tensorial notation, as follows: 
\bea
&&
\kappa \{ g_1, g_2\}_P = g_1 g_2 \hat r(M^I) - 
\hat r(M^F) g_1 g_2 
\nonumber\\
&& \kappa \{ g_1, M^I_2\}_P = g_1 M_2^I \hat \Theta(M^I)
\nonumber\\
&& \kappa \{ g_1, M_2^F\}_P = M_2^F \hat\Theta(M^F) g_1
\nonumber\\
&&\kappa \{ M^I_1, M^I_2\}_P = M^I_1 M^I_2 \hat\Delta(M^I)
\nonumber\\
&&\kappa \{ M^F_1, M^F_2\}_P = - M^F_1 M^F_2 \hat\Delta(M^F)
\nonumber\\
&&\kappa \{ M^I_1, M^F_2\}_P =0.
\label{groupoidPB}
\eea
Here $\kappa$ is an arbitrary constant included for comparison purposes.
The `structure functions' $\hat r$, $\hat \Theta$, $\hat \Delta$ 
are $\G\otimes \G$ valued functions 
on $\check G$; in components 
\be
\hat r(M)= r^{\alpha\beta}(M) T_\alpha\otimes T_\beta,
\quad
\hat \Theta(M)= \Theta^{\alpha\beta}(M) T_\alpha\otimes T_\beta,
\quad
\hat \Delta(M)= \Delta^{\alpha\beta}(M) T_\alpha\otimes T_\beta.
\ee 
It is quite easy to verify that a PB given by
the ansatz (\ref{groupoidPB}) 
always yields a PL groupoid, since the constraints
in (\ref{graph}) will be first class for any choice
of the structure functions.
Of course, the structure functions must satisfy a system
of equations in order for the above ansatz to define a PB.
The antisymmetry of the PB is ensured by
\be 
\hat r=-\hat r_{21}
\quad (\hat r_{21}:= r^{\alpha\beta} T_\beta \otimes T_\alpha)
\qquad\hbox{and}\qquad \hat 
\Delta=-\hat \Delta_{21},
\ee
while  the Jacobi identity is, in fact, equivalent to the following
system of equations:
\bea
&& 
[ \hat r_{12}, \hat r_{13}] + \Theta_{\alpha\beta}
T^\alpha_1 {\cal R}^\beta \hat r_{23} + \hbox{cycl. perm.}= \mu \hat f ,
\quad \mu=\hbox{constant},
\label{J1}\\
&& 
[ \hat \Delta_{12}, \hat \Delta_{13}] + 
\Delta_{\alpha\beta} T^\alpha_1 {\cal R}^\beta \hat \Delta_{23}
+ \hbox{cycl. perm.}=0, 
\label{J4}\\
&& [\hat r_{12}, \hat \Theta_{13}+\hat \Theta_{23}] + 
[\hat \Theta_{13}, \hat \Theta_{23}] 
+\Delta_{\alpha\beta} T^\alpha_3 {\cal R}^\beta \hat r_{12} 
+\Theta_{\alpha\beta}( T^\alpha_1 {\cal R}^\beta \hat \Theta_{23}
- T^\alpha_2 {\cal R}^\beta \hat \Theta_{13})=0,\qquad\qquad
\label{J2}\\
&& [\hat\Theta_{12}+\hat\Theta_{13}, \hat\Delta_{23}] +
[\hat\Theta_{12},\hat \Theta_{13}]
+\Theta_{\alpha\beta} T^\alpha_1 {\cal R}^\beta \hat\Delta_{23}
+\Delta_{\alpha\beta} (T^\alpha_3 {\cal R}^\beta \hat \Theta_{12} 
-T^\alpha_2 {\cal R}^\beta \hat\Theta_{13})=0.\qquad
\label{J3}\eea
Observe that the left hand side of (\ref{J1}) 
is of the same form as that of (\ref{GCDYB}), but in the groupoid context
on the right hand side we have an {\em arbitrary} constant $\mu$.
The derivation of the above equations from the various instances of 
the Jacobi identity is not difficult. What is somewhat miraculous 
is that one does not obtain more equations than these.
This is actually ensured by our choice of the relationship between
the PBs that involve $M^I$ and those that involve $M^F$.
As an illustration, let us explain how 
(\ref{J1}) is derived.
By evaluating 
\be
\{ \{g_1, g_2\}_P, g_3\}_P + \hbox{cycl. perm.}=0,
\ee
one obtains that this is equivalent to 
\bea
&&g_1 g_2 g_3 
\left([ \hat r_{12}, \hat r_{13}] + \Theta_{\alpha\beta}
T^\alpha_1 {\cal R}^\beta \hat r_{23} + \hbox{cycl. perm.}\right)(M^I)=
\nonumber \\
&&\qquad =
\left([ \hat r_{12}, \hat r_{13}] + \Theta_{\alpha\beta}
T^\alpha_1 {\cal R}^\beta \hat r_{23} + 
\hbox{cycl. perm.}\right)(M^F) g_1 g_2 g_3.
\eea
This holds if and only if the expression in the parenthesis
is a constant, ${\mathrm Ad}$-invariant element of $\wedge^3(\G)$,
and $\mu \hat f$ is the only such element for a simple Lie algebra $\G$.

We have seen that the chiral WZNW  
PBs are encoded by equations (\ref{xchPB}), (\ref{Omega1}) and
(\ref{Omegabar1}), where $\hat \Theta$ and $\hat \Delta$ are defined 
by (\ref{Theta}) and (\ref{Delta}) respectively  
in terms of a solution $\hat r$ of (\ref{GCDYB}).
Now our point is the following: {\em A PL groupoid can be 
naturally associated 
with any Poisson structure on the chiral WZNW phase space 
by taking the triple $\hat r$, $\hat \Theta$, $\hat \Delta$
that arises in the WZNW model to be the structure functions of a 
PL groupoid 
according to (\ref{groupoidPB}).}

It can be checked that the Jacobi identities of the PL groupoid
(\ref{J1})--(\ref{J3}) are satisfied for any triple 
$\hat r$, $\hat \Theta$, $\hat \Delta$ that arises in the WZNW model. 
This actually follows without any computation since, indeed, 
the Jacobi identities of the chiral WZNW PBs in (\ref{xchPB}), 
(\ref{Omega1}),
(\ref{Omegabar1}) lead to the same equations, with $\mu=-{1\over 4}$, 
and they are satisfied since they follow from the 
symplectic form $\kappa \Omega_{chir}^\rho$.   

Among the `chiral WZNW PL groupoids' described above there are 
those special cases for which $\hat K= \hat r- \hat R$ satisfies
the equivariance condition  
(\ref{equi2}) in relation with some constant r-matrix 
$\hat R$ subject to (\ref{CDYBconst}).
In these cases,  
we equip the group $G=\{ h\}$ with the Sklyanin bracket
opposite to that in (\ref{PBgroup}),
\be
\big\{h\stackrel{\otimes}{,}h\big\}={1\over\kappa}\,
\Big[h\otimes h, \hat R\Big],
\label{-PBgroup}\ee
and consider its commuting right and left actions on $P$ 
given respectively by the maps 
\be
P \times G \ni \bigl( (M^F, g, M^I), h \bigr) \mapsto (M^F, gh, 
h^{-1} M^I h)\in P
\label{rightaction}\ee
and 
\be
G \times P\ni \bigl( h, (M^F, g, M^I)\bigr) \mapsto (hM^F h^{-1},
hg, M^I)\in P.
\label{leftaction}\ee
Then it is not difficult to verify that these are {\em both} Poisson maps,
i.e., they define  two PL 
actions
of the PL group $G$ (with (\ref{-PBgroup}))
on the PL groupoid $P$.
In the final analysis, 
this is a consequence of the fact that, as explained in sec.~5,
in the present situation  we have a PL action 
of $G$ on the chiral WZNW phase space whose 
Poisson structure is encoded by $(P, \{\ ,\ \}_P)$.
Here $\check G\subset G$ must be $\Ad$-invariant, see footnote 5. 
 
In \cite{EV} PL groupoids are associated with arbitrary subalgebras 
${\cal K}\subset \G$, although the corresponding dynamical r-matrices 
are described only if ${\cal K}$ is a Cartan subalgebra.
The ${\cal K}=\G$ special case of their 
groupoids  is in fact equivalent to our PL groupoid whose structure 
function is the r-matrix in (\ref{ralt}).
Their PL groupoids are different from ours in general.

\section{Conclusion}

In this paper we explored  
the Poisson structures on the chiral WZNW phase space of 
group valued quasiperiodic fields with generic monodromy.
We have shown that the possible PBs are defined by 
the exchange r-matrices that are solutions of  (\ref{GCDYB}).
This equation can be viewed as an  analogue of the 
celebrated CDYB equation (\ref{CDYB}),
since the latter plays a similar r\^ole for chiral WZNW fields 
with diagonal monodromy.
An analysis of chiral WZNW Bloch waves and their classical Wakimoto 
realizations
in the spirit of the present paper is contained  in \cite{BFP2}. 

We have given an interpretation of our dynamical YB 
equation (\ref{GCDYB}) in terms of 
a family of PL groupoids, whose further study may be fruitful.
In this respect, the most interesting open questions appear to be
to quantize these PL groupoids and to find applications for them
outside the chiral WZNW context.
It is known that equation (\ref{CDYB}) admits interesting applications 
in the  field of integrable systems \cite{Avan}.

We also investigated the PL symmetries of the exchange algebra (\ref{xchPB}).
We have found that for {\em any} PL structure on
the WZNW group $G$ there is a corresponding choice of the 
exchange r-matrix such
that the standard  gauge action of $G$ on the chiral fields becomes a
PL action.
It would be desirable to understand if this result has any analogue 
at the level of the quantized (chiral) WZNW model.

\bigskip
\bigskip
\noindent
{\bf Acknowledgements.}
This investigation was supported in part by the Hungarian National
Science Fund (OTKA) under T019917, T030099, T025120 
and by the Ministry
of Education under FKFP 0178/1999.

\renewcommand{\thesection}{\Alph{section}}
\setcounter{section}{0} 

\section{Exchange r-matrices for $SU(2)$}
\setcounter{equation}{0}
\renewcommand{\theequation}{A.\arabic{equation}}

In this appendix we present 
an explicit, local formula
for the most general exchange r-matrix on the simplest compact 
Lie group $G=SU(2)$.
The formula (\ref{Rtelj}) below is valid in a neighbourhood of the 
unit element. 
It illustrates some general results obtained in sec.~3,
and it may prove useful in a future study of the question whether globally 
defined exchange r-matrices exist for $SU(2)$ or not.

As a basis for the Lie algebra $su(2)$, we choose the generators
$T^a:= {1\over 2i} \sigma_a$, where the $\sigma_a$ ($a=1,2,3$) are 
the usual Pauli matrices.
We parametrize the matrices $r^{ab}(M)$ and $q^{ab}(M)$ 
that appear in (\ref{requfact}) in terms of 3-component vectors as
\be\label{vecform}
r^{ab}=\epsilon^{abc}R_c\qquad\hbox{and}\qquad q^{ab}=\epsilon^{abc}Q_c,
\ee
where $\epsilon^{abc}$ is the totally antisymmetric tensor for which 
$\epsilon^{123}=1$. 
Furthermore, we identify the $SU(2)$ group manifold with $S^3\subset \R^4$
by writing $M\in SU(2)$ according to  
\be\label{xsystem}
M=x_0 \sigma_0 +ix_a\sigma_a\,,\qquad\qquad x_0^2+x_a x_a=1,
\ee
whereby $x_0$, $x_a$ define smooth functions  on $SU(2)$
($\sigma_0$ is the $2\times 2$ unit matrix).
It is then a matter of straightforward  calculation to translate 
eq.~(\ref{rexplicit}) into the formula
\be\label{RfromQ}
R_a={1\over 4} D^{-1}\left(x_a+4x_0Q_a+4\epsilon^{abc}x_bQ_c\right),
\qquad  
D :=x_0-4x_a Q_a,
\ee
which is valid on a neighbourhood of the unit element 
where $D \neq 0$. 
By assumption, on this neighbourhood the
$Q_a$ are smooth functions subject to 
\be
d\rho(M)={1\over 6} 
{\mathrm tr} (M^{-1} dM)^{3\wedge}
\quad\hbox{for}\quad  
\rho(M)= {1\over 2}\,q^{ab}(M){\mathrm{tr}}
\big(T_a M^{-1}dM\big)\wedge{\mathrm{tr}}\big(T_b M^{-1}dM\big).
\label{Qeq}\ee
As discussed in sec.~3.3,  (\ref{Qeq}) implies that  
$R_a$ defined by (\ref{vecform}), (\ref{RfromQ})
yields a solution of the dynamical YB equation (\ref{GCDYB}),
which for $SU(2)$ can actually be written in the form
\be
2 R_a R_a + {1\over 2} {\cal D}^{a+} R_a +\epsilon^{abc} R_c 
{\cal D}^-_b R_a = -{1\over 8}.
\label{Req}\ee    
Observe that we cannot have a constant solution
since the $R_a$ must be real. 
It is also worth noting that locally we have the 
inverse of (\ref{RfromQ}) given by
\be\label{QfromR}
Q_a = {1\over 4} {\tilde D}^{-1} \left(-x_a + 4 x_0 R_a +
4 \epsilon^{abc} x_b R_c\right),
\qquad
\tilde D :=x_0 + 4x_a R_a.
\ee
This formula defines via (\ref{vecform}) a solution of (\ref{Qeq})
out of any solution of (\ref{Req}).

We shall now derive the general local solution of (\ref{Req})
by making explicit the general local solution (\ref{beta}) of (\ref{Qeq})
that we have obtained in sec.~3.5 for any group.
For this we need the exponential parametrization of $SU(2)$, 
\be
M=e^{2\pi\Gamma}=e^{2\pi\Gamma_aT^a}
\qquad\hbox{with}\qquad
\vert \Gamma \vert^2 = \Gamma_a \Gamma_a <1,
\ee
which covers the domain $SU(2)\setminus \{ - \sigma_0\}$.
Upon comparison with (\ref{xsystem}), we get   
\be 
x_0=\cos(\pi \vert\Gamma\vert),\qquad 
x_a=-\frac{\Gamma_a}{\vert\Gamma\vert}\sin(\pi
\vert\Gamma\vert).
\ee
We shall also use the following expressions  
for the powers of the operator $\ad \Gamma$. 
For the odd powers,  we have 
\be
(\ad \Gamma)^{2n+1} =(-1)^n\vert\Gamma\vert^{2n} (\ad \Gamma),
\quad n\ge 0,
\qquad
(\ad \Gamma) (T^a) = [\Gamma, T^a]= \epsilon^{abc}  T_b  \Gamma_c.
\ee
For the even powers, 
defining the matrix of $(\ad \Gamma)^n$ by 
$(\ad \Gamma)^n (T^b)= [(\ad \Gamma)^n ]_{a}^{\phantom{a} b} T^a$, we have    
\be
[(\ad \Gamma)^{2n}]_a^{\phantom{a}b}=(-1)^n\vert\Gamma\vert^{2n}\,\Bigl(
\delta_{ab} -\frac{\Gamma_a\Gamma_b }{\vert\Gamma\vert^2}\Bigr), 
\quad n\ge 1.
\ee
Using these relations, we can rewrite the formulae (\ref{qexpl}) 
and (\ref{ralt}) as follows:
\be\label{q0}
q_0^{ab}= \epsilon^{abc} Q^{(0)}_c  
\quad\hbox{with}\quad
Q^{(0)}_c=\Gamma_c\frac{2\pi \vert\Gamma\vert -
\sin(2\pi \vert\Gamma\vert)}{8\vert\Gamma\vert\sin^2(\pi \vert\Gamma\vert)}\,,
\ee
and
\be\label{su2rm}
r_0^{ab} = \epsilon^{abc} R^{(0)}_c 
\quad\hbox{with}\quad
R^{(0)}_c =
\frac{1}{4}\frac{\Gamma_c}{\vert\Gamma\vert}\Big(\cot(\pi \vert\Gamma\vert)
-\frac{1}{\pi \vert\Gamma\vert}\Big)\,.
\ee
One may check that eqs.~(\ref{RfromQ})--(\ref{QfromR}) hold 
for these expressions, which represent smooth functions 
on $SU(2)\setminus \{ - \sigma_0\}$.

To obtain the most general 2-form $\rho$ on $SU(2)\setminus \{ - \sigma_0\}$
that satisfies (\ref{Qeq}), we have to add an arbitrary closed
2-form to the 2-form, $\rho_0$, that corresponds to the matrix $q_0^{ab}$.
In fact, the result can be written as 
\be
\rho(\Gamma )=d\Gamma_a\wedge
d\Gamma_b \epsilon^{cba} \big(\Gamma_c\frac{\sin (2\pi \vert\Gamma\vert
  )-2\pi \vert\Gamma\vert}{2\vert\Gamma\vert^3} + U_c(\Gamma)\big)\,,
\label{rhoU}\ee
where $U_a(\Gamma)$ is a smooth `vector-function' in the interior 
of the unit ball, $\vert\Gamma\vert <1$, which is divergence free, i.e.,
\be
\sum_{a=1}^3 {\partial U_a(\Gamma) \over \partial \Gamma_a}=0.
\label{divU}\ee
We then have to rewrite this 2-form in the manner indicated by 
the second parts of (\ref{vecform}) and (\ref{Qeq}).
By means of (\ref{RfromQ}), this will provide us with the general local
solution of (\ref{Req}). 
By performing the necessary (rather tedious) calculations,
in this way we obtain the following formula:  
\be
R_a=\frac{\Gamma_a}{\vert\Gamma\vert}
\left[ \frac{1}{4} \cot(\pi\vert\Gamma\vert )
-\frac{1}{\vert\Gamma\vert (4\pi-2 \Gamma\cdot U )}\right]
+\left[ \frac{\Gamma_a}{\vert\Gamma\vert}(\Gamma \cdot U )-
\vert\Gamma\vert U_a
\right]\frac{1}{(4\pi-2 \Gamma \cdot U)2\pi \vert\Gamma\vert},
\label{Rtelj}
\ee
where $\Gamma \cdot U=\Gamma_a U_a$. 
This expression is valid on the open subset of $SU(2)$ 
that excludes $-\sigma_0$ and the points where $\Gamma \cdot U=2\pi$.
In particular, $R_a$ is smooth in a neighbourhood 
of the unit element, for which $\Gamma =0$, 
since $U_a(\Gamma)$ is smooth there by assumption.
One can also verify explicitly that on its domain of validity 
$R_a$  solves the dynamical YB equation (\ref{Req}) for
any divergence free  $U_a(\Gamma)$.
For this verification, one needs to spell out (\ref{Req}) 
more explicitly. For instance, if one uses the $x_a$ in (\ref{xsystem})
as coordinates around $\sigma_0\in SU(2)$, then (\ref{Req}) becomes
\be
2R_a R_a-{\sqrt{1- x_b x_b}\over 2} {\partial R_a\over\partial x_a} 
+2x_a R_b {\partial R_a\over\partial x_b} -
2R_b x_b {\partial R_a\over\partial x_a} =-{1 \over 8}.
\label{GCDYB3}
\ee 

In summary, we have derived the form of the most general 
exchange r-matrix in a neighbourhood of the unit element of $SU(2)$. 
The solution (\ref{Rtelj}) may in general develop singularities 
away from the unit element, and it is 
an open question if globally smooth  solutions of (\ref{Req}) 
exist on $SU(2)$ or not. 

\section{Analytic solution of the dynamical YB equation} 
\setcounter{equation}{0}
\renewcommand{\theequation}{B.\arabic{equation}}

In this appendix we show that the equivariant analytic function 
(\ref{soln}) is a solution of the dynamical YB equation (\ref{dyneq}).

We will use the power series expansion of the $\coth$ function:
\be
\coth z={1\over z}+\sum_{r=0}^\infty\,\alpha_r z^{2r+1}\,.
\label{coth}
\ee
Here the coefficients $\alpha_r$ can be expressed in terms of the
Bernoulli numbers \cite{GR}. They can also be computed using the recursion
relation
\be
\alpha_0={1\over3}\,,\qquad\qquad
\alpha_m=-{1\over2m+3}\sum_{r=0}^{m-1}\,\alpha_r\alpha_{m-1-r}\,,
\qquad\qquad m=1,2,\dots
\label{recursion}
\ee

Using the properties of the operator ${\cal D}^{+\alpha}$ (\ref{Dpm})
and the definition of $K$ (\ref{soln}) the first two terms of (\ref{dyneq})
can be expanded as
\be
(K{\T}^\alpha K)^{\beta\gamma}=\sum_{r,s=0}^\infty\,
\alpha_r\alpha_s
\Big({1\over2^{2r+2}}-\nu^{2r+2}\Big)\,
\Big({1\over2^{2s+2}}-\nu^{2s+2}\Big)\,
\Big({\cal Y}^{2s+1}{\T}^{\alpha}{\cal Y}^{2r+1}\Big)^{\beta\gamma}
\label{first}
\ee
and
\be
{1\over2}{\cal D}^{+\alpha}K^{\beta\gamma}=\sum_{r=0}^\infty\,
\alpha_r\Big({1\over2^{2r+2}}-\nu^{2r+2}\Big)\,\sum_{s=0}^{2r}\,
\Big({{\cal Y}\over2}\coth{{\cal Y}\over2}\Big)^{\alpha\lambda}
\Big({\cal Y}^s{\T}_\lambda{\cal Y}^{2r-s}\Big)^{\beta\gamma}\,.
\label{second}
\ee
Here, in writing the second equality we exploited that in the
parametrization of $M$ introduced in sec.~3.5 one readily obtains 
by writing ${\cal Y}={\cal X}^a{\cal T}_a$ that
\be
{\cal L}^\alpha ({\cal X}^\beta)=
\Bigl(\frac{{\cal Y}}{1-\exp(-{\cal Y})}\exp(-{\cal
  Y})\Bigr)^{\alpha\beta},\qquad
  {\cal R}^\alpha ({\cal X}^\beta)=
\Bigl(\frac{{\cal Y}}{\exp({\cal Y})-1}\exp({\cal
  Y})\Bigr)^{\alpha\beta}\,.
\ee
It is clear that all terms in (\ref{dyneq}) are built from 
powers of ${\cal Y}$ and structure constants. It will
prove useful to contract all the indices
with Lie algebra generators and thus reformulate (\ref{dyneq}) as an equation
in the triple tensor product of the Lie algebra. We introduce the notation
\be
\langle k,l,m\rangle:=
\big({\cal Y}^k\big)^{\alpha\kappa}
\big({\cal Y}^l\big)^{\beta\lambda}
\big({\cal Y}^m\big)^{\gamma\sigma}
f_{\kappa\lambda\sigma}\,T_\alpha\otimes T_\beta\otimes T_\gamma\,.
\label{notation}
\ee
Cyclic permutation of the indices now corresponds to cyclic permutation
of the tensor factors and we also introduce the  symbol
\be
[k,l,m]:=
\langle k,l,m\rangle +
\langle l,m,k\rangle +
\langle m,k,l\rangle\,.
\label{cycl}
\ee

We expand both sides of (\ref{dyneq}) in powers of $\nu$ and every
coefficient of $\nu$ 
also in powers of ${\cal Y}$. In our notation this 
latter expansion corresponds to putting together all terms with a fixed
total $N=k+l+m$, and eq.~(\ref{dyneq}) requires separately the equality
of all such terms on the two sides. 

We start with the $\nu^0$ terms. We find that the $N=0$ piece 
is satisfied identically, while for $N=2m+2$ ($m=0,1,\dots$) we get
\bea
&&\alpha_{m+1}\sum_{s=0}^{2m+2}\,(-1)^{s+1}[ 0,s,2m+2-s]
-\sum_{r+s=m}\alpha_r\alpha_s[ 0,2s+1,2r+1]\nonumber\\
&&+\sum_{r+k=m}\alpha_r\alpha_k\sum_{s=0}^{2r}(-1)^{s+1}[
2k+2,s,2r-s]=0\,.\label{nu0}
\eea
Similarly the $N=0$ piece of the $\nu^2$ term vanishes identically, while
for $N=2r+2$ ($r=0,1,\dots$) we obtain 
\be
[ 2r+2,0,0]+
[ 0,1,2r+1]+
[ 0,2r+1,1]=0\,.
\label{nu2}
\ee

Finally, from the $\nu^{2m+4}$ terms (for $m=0,1,\dots$) we get
contributions with $N=(2m+2)$-th powers of ${\cal Y}$,
\be
\alpha_{m+1}\sum_{s=0}^{2m+2}(-1)^s[0,s,2m+2-s]
-\sum_{r+s=m}\alpha_r\alpha_s[0,2s+1,2r+1]=0,
\label{num}
\ee
and also terms with $N=(2m+4+2r)$-th powers of ${\cal Y}$ (for $r=0,1,\dots$),
\be
[0,2m+3,2r+1]+[0,2r+1,2m+3]+
\sum_{s=0}^{2m+2}(-1)^s[2r+2,s,2m+2-s]=0\,.
\label{numr}
\ee

Before proceeding we note that using the Jacobi identity for the
structure constants we can write down the following identity:
\be
\langle k+1,l,m\rangle+
\langle k,l+1,m\rangle+
\langle k,l,m+1\rangle=0\,.
\label{shift1}
\ee

Now it is easy to see that (\ref{nu2}) is a special case and
(\ref{numr}) is a simple consequence of the above identity. In fact, to 
prove (\ref{numr}) we group $[0,2m+3,2r+1]$ with the first
($[0,2r+1,2m+3]$ with the last) $m+1$ terms of the sum:
\bea
&&Z_1=[0,2m+3,2r+1]+\sum_{s=0}^{m}(-1)^s[2r+2,s,2m+2-s],\nonumber\\
&&Z_2=[0,2r+1,2m+3]+\sum_{j=0}^{m}(-1)^j[2r+2,2m+2-j,j]\,,\nonumber
\eea
 then in both
groups, for odd $s$ ($j$) we use (\ref{shift1}) to write
\bea
&&(-1)^s[2r+2,s,2m+2-s]=[2r+1,s+1,2m+2-s]+[2r+1,s,2m+3-s],\nonumber\\
&&(-1)^j[2r+2,2m+2-j,j]=[2r+1,2m+3-j,j]+[2r+1,2m+2-j,j+1].\nonumber
\eea
In this form, as a consequence of (\ref{shift1}), both $Z_1$ and $Z_2$
cancel \lq telescopically' almost completely; for odd $m$ they give
\be
Z_1\mapsto [2r+1,m+1,m+2],\qquad Z_2\mapsto [2r+1,m+2,m+1],
\ee
while for even $m$ we get
\bea
&&Z_1\mapsto [2r+2,m,m+2]+[2r+1,m,m+3],\nonumber\\
&&Z_2\mapsto [2r+2,m+2,m]+[2r+1,m+3,m]\,.
\eea
One readily verifies, using again 
 (\ref{shift1}), that these remaining terms 
give zero in both cases with the \lq central' element of the sum 
$(-1)^{m+1}[2r+2,m+1,m+1]$.
   
As for (\ref{nu0}), it should be perfectly possible to prove it using 
the properties of the coefficients $\alpha_r$, but in the present
context there is no need to prove it independently since we know
that the $\nu=0$ case coincides with the solution (\ref{ralt})
and therefore it must satisfy (\ref{GCDYB}).

Thus we are left with (\ref{num}). Using (\ref{nu0}), we can
rewrite it as
\be
2\sum_{r+k=m}\alpha_r\alpha_k[ 0,2r+1,2k+1]
+\sum_{r+k=m}\alpha_r\alpha_k\sum_{s=0}^{2r}(-1)^s[
2k+2,s,2r-s]=0
\ee
and in this form we see that it trivially follows from
(\ref{numr}) and the index symmetry of $\alpha_k\alpha_r$.

\end{document}